\documentclass[9pt,shortpaper,twoside,web]{ieeecolor}
\usepackage{generic}
\usepackage[english]{babel}
\usepackage[bookmarks=false]{hyperref}
\addto\extrasenglish{

}
\usepackage{amsmath,amssymb,amsfonts}
\usepackage{algorithmic}
\usepackage{graphicx}
\usepackage[normalem]{ulem}
\usepackage{textcomp}
\usepackage{multirow}
\usepackage{amsmath}
\usepackage[ruled,vlined]{algorithm2e}
\usepackage{setspace}
\usepackage{float}
\usepackage[]{caption}
\usepackage{tabularx}
\usepackage{cite}
\usepackage[font=small]{caption}
\usepackage{threeparttable}
\def\BibTeX{{\rm B\kern-.05em{\sc i\kern-.025em b}\kern-.08em
    T\kern-.1667em\lower.7ex\hbox{E}\kern-.125emX}}
\begin{document}
\bstctlcite{IEEEexample:BSTcontrol}

\title{MetaSleepLearner: A Pilot Study on Fast Adaptation of Bio-signals-Based Sleep Stage Classifier to New Individual Subject Using Meta-Learning}

\author{Nannapas~Banluesombatkul, Pichayoot~Ouppaphan, Pitshaporn~Leelaarporn, Payongkit~Lakhan, Busarakum~Chaitusaney, Nattapong~Jaimchariyatam, Ekapol~Chuangsuwanich, \\Wei Chen, \IEEEmembership{Senior Member, IEEE}, Huy~Phan, \IEEEmembership{Member, IEEE} Nat~Dilokthanakul$^{*}$ and Theerawit~Wilaiprasitporn$^{*}$, \IEEEmembership{Member, IEEE}
\thanks{This work was supported by PTT Public Company Limited, The SCB Public Company Limited, Thailand Science Research and Innovation (SRI62W1501) and Office of National Higher Education Science Research and Innovation Policy
Council (C10F630057).}
\thanks{N. Banluesombatkul, P. Ouppaphan, P. Leelaarporn, P. Lakhan, N. Dilokthanakul and T. Wilaiprasitporn are parts of Bio-inspired Robotics and Neural Engineering Lab, School of Information Science and Technology, Vidyasirimedhi Institute of Science \& Technology, Rayong, Thailand {\tt\small ($^{*}$corresponding authors: natd\_pro at vistec.ac.th, theerawit.w at vistec.ac.th)}.}
\thanks{B. Chaitusaney is with Department of Otolaryngology, Faculty of Medicine, Chulalongkorn University. She is also with Excellence Center for Sleep Disorders, King Chulalongkorn Memorial Hospital, Thai Red Cross Society, Bangkok, Thailand.}
\thanks{N. Jaimchariyatam is with Division of Pulmonary and Critical Care Medicine, Faculty of Medicine, Chulalongkorn University. He is also Head of Excellence Center for Sleep Disorders, King Chulalongkorn Memorial Hospital, Thai Red Cross Society, Bangkok, Thailand.}
\thanks{E. Chuangsuwanich is with the Computer Engineering Department, Chulalongkorn University, Bangkok, Thailand.}
\thanks{W. Chen is with the Center for Intelligent Medical Electronics, Department of Electronic Engineering, School of Information Science and Technology, Fudan University, Shanghai 200433, China}
\thanks{H. Phan is with the School of Electronic Engineering and Computer Science, Queen Mary University of London, United Kingdom.}
}

\maketitle
\begin{abstract}
Identifying bio-signals based-sleep stages requires time-consuming and tedious labor of skilled clinicians. Deep learning approaches have been introduced in order to challenge the automatic sleep stage classification conundrum. However, the difficulties can be posed in replacing the clinicians with the automatic system due to the differences in many aspects found in individual bio-signals, causing the inconsistency in the performance of the model on every incoming individual. Thus, we aim to explore the feasibility of using a novel approach, capable of assisting the clinicians and lessening the workload. We propose the transfer learning framework, entitled MetaSleepLearner, based on Model Agnostic Meta-Learning (MAML), in order to transfer the acquired sleep staging knowledge from a large dataset to new individual subjects (source code is available at https://github.com/IoBT-VISTEC/MetaSleepLearner). The framework was demonstrated to require the labelling of only a few sleep epochs by the clinicians and allow the remainder to be handled by the system. Layer-wise Relevance Propagation (LRP) was also applied to understand the learning course of our approach. In all acquired datasets, in comparison to the conventional approach, MetaSleepLearner achieved a range of 5.4\% to 17.7\% improvement with statistical difference in the mean of both approaches. The illustration of the model interpretation after the adaptation to each subject also confirmed that the performance was directed towards reasonable learning. MetaSleepLearner outperformed the conventional approaches as a result from the fine-tuning using the recordings of both healthy subjects and patients. This is the first work that investigated a non-conventional pre-training method, MAML, resulting in a possibility for human-machine collaboration in sleep stage classification and easing the burden of the clinicians in labelling the sleep stages through only several epochs rather than an entire recording.
\end{abstract}

\begin{IEEEkeywords}
Sleep stage classification, meta-learning, pre-trained EEG, transfer learning, convolutional neural network.
\end{IEEEkeywords}

\section{Introduction}
\label{sec:introduction}
\iffalse
\IEEEPARstart{I}{n} the present day, the technology revolving around healthcare engages in the importance of the quality of sleep for better understanding of different sleep-related disorders such as insomnia and obstructive sleep apnea (OSA) \cite{rundo2019polysomnography, kang2017state}. Furthermore, a fundamental knowledge of sleep staging can be applied to some other novel purposes such as the measuring of an effectiveness of vagus nerve stimulation (VNS) therapy in patients with epilepsy \cite{ravan2019investigating} and a development of wearable sleep monitoring devices using in-ear electroencephalography (EEG) \cite{nakamura2019hearables} or other consumer-grade EEG devices \cite{sawangjai2020consumer}.
\fi
\IEEEPARstart{A} conventional method of measuring the sleep stages is to conduct the gold standard sleep study called polysomnography (PSG) in sleep laboratory or medical facility \cite{rundo2019polysomnography, malhotra2018polysomnography}. Various sensing techniques, such as electroencephalography (EEG), electro-oculography (EOG), sub-mental electromyography (EMG), electrocardiography (ECG), airflow, etc., are combined to detect the electrical signals emitted from different parts of the human body \cite{boostani2017comparative, faust2019review}. During the recording, each 30-second segment of signal interval, i.e., epoch, is annotated by sleep experts to either one of seven stages (Wake (W), Non-Rapid Eye Movement (NREM: S1, S2, S3, and S4), Rapid Eye Movement (REM), and movement time (MT)) following the Rechtschaffen and Kales (R\&K) rules \cite{moser2009sleep}, or one of five stages (W, REM, and NREM: N1-N3) according to the more recently presented procedure from the American Academy of Sleep Medicine (AASM) \cite{berry2012aasm}. 

However, the manual scoring, throughout a whole night of sleeping test (approximately 8 hours or 1000 epochs), is time-consuming and burdening the human labor. Therefore, numerous studies have been proposing automatic sleep stage classification by using the main signals containing the characteristics of each stage including EEG, EOG, and sub-mental EMG. The techniques can be categorized into three groups: 1) human-engineered feature extraction with automatic decision making algorithms \cite{chen2015symbolic, aboalayon2016sleep, li2017hyclasss, HASSAN2017201, HASSAN201776, HASSAN2017115, gharbali2018investigating, alickovic2018ensemble, JIANG2019HMM, plausibility2020zhang}, 2) the application of extracted features on the Deep Learning (DL) approach \cite{tsinalis2016automatic, hierarchical2020chen}, and 3) the use of an end-to-end training of the DL approach including Convolutional Neural Network (CNN) or Recurrent Neural Network (RNN) \cite{dong2017mixed, chambon2018deep, sors2018convolutional, korkalainen2019accurate, perslev2019utime, phan2019, wei2020residual}, which involved state-of-the-art sleep staging networks, e.g., DeepSleepNet \cite{supratak2017deepsleepnet} and SeqSleepNet \cite{phan2019seqsleepnet}.

% Different human-engineered feature extraction methods, based on the knowledge of sleep staging in the field of sleep medicine, have been developed before feeding them into several decision making techniques such as symbolic fusion \cite{chen2015symbolic} or Machine Learning for sleep stage classification \cite{aboalayon2016sleep, li2017hyclasss, gharbali2018investigating, alickovic2018ensemble, JIANG2019HMM, plausibility2020zhang}. Although, using those features allows us to understand the model easily, Deep Learning (DL) approach has been implemented to achieved higher performance. Due to the capability of either Convolutional Neural Network (CNN) which can extract the features automatically without the requirement of the domain knowledge or the Recurrent Neural Network (RNN) which has a capability to learn temporal context of time-series data \cite{dong2017mixed, chambon2018deep, sors2018convolutional, phan2019}, it results the state-of-the-art sleep staging networks, e.g. DeepSleepNet \cite{supratak2017deepsleepnet} and SeqSleepNet \cite{phan2019seqsleepnet}, as well as other studies which achieved comparable performance but better in other aspects i.e. lower training time \cite{wei2020residual}. Moreover, to avoid the fully blackbox of DL networks, recently publish works from Sun \textit{et al.} fuses the knowledge based features with the ones from CNN \cite{hierarchical2020chen} and achieved similar performance.

Despite the ability to achieve high accuracy of DL model, the method is still not practical due to the massive amount of data needed for training. Furthermore, the models trained on one cohort are not directly applicable to another one due to the data variation from a number of reasons in practice including the amount and the placement of EEG channels, sampling frequencies, experimental protocols, and types of subjects \cite{boostani2017comparative, andreotti2018multichannel}, limiting its applicability in clinical environments.

% Therefore, it is difficult to ensure the achieved high accuracy of the selected models from newly established datasets, especially in the small cohorts.

% Despite the application of the state-of-the-art sleep stage classification models with the capability of DL, numerous amounts of data are required in order to achieve high performance. However, in a real-world situation, each cohort starts from having only a few samples and may have used different recording systems. The number and the placement of EEG channels, sampling frequencies, experimental protocols, and types of subjects may lead to data variation and the alteration in the performance of the classification model \cite{boostani2017comparative, andreotti2018multichannel}. Therefore, it is difficult to ensure the achieved high accuracy of the selected models from newly established datasets or devices.

According to aforementioned limitations of the existing DL approaches, one solution for the limited number of samples problem is \textit{transfer learning (TL)} methodology. TL paradigm has a two-step training procedure. The model is pre-trained by a huge dataset, followed by the application (fine-tuning) to the new in-coming dataset \cite{andreotti2018multichannel}. Further elaborations are described in \autoref{sec:TL}.

%In our previous work, we have shown that the TL methodology is possible to apply to EEG data in ERP classification task \cite{PretrainP300}. 
% \textcolor{red}{The performance of event-related potential (ERP) classification tasks could be improved by pre-training the model from other datasets and fine-tune to the target one.}
%Recently, researchers have also begun to employ TL method for sleep stage classification. Some methods were applied directly from the image classification models \cite{vilamala2017deep} while many of them transferred sleep staging knowledge from large sleep cohorts to other smaller cohorts \cite{phan2019deep, phan2019towards}. \textcolor{red}{However, signals recorded from each subject has a high variation e.g. the amplitude, sleep pattern, etc. ?? \cite{buckelmuller2006trait}. Therefore, the model performs with different accuracy level for different subjects. Thus, fine-tuning is needed to ensure the reliability of the model.}

In our previous work, we have shown that it is possible to apply the TL methodology, i.e., pre-training the Auto-encoder, to EEG data in an event-related potential (ERP) classification task \cite{PretrainP300}. Recently, researchers have also begun to employ TL method for sleep stage classification. Some methods were applied directly from the image classification models \cite{vilamala2017deep}, while many of them transferred sleep staging knowledge from large sleep cohorts to other smaller cohorts \cite{phan2019deep, phan2019towards}. However, signals recorded from each subject has a high variation, e.g., sleep pattern, sleep stage duration, and EEG power spectra, etc. \cite{buckelmuller2006trait}. Therefore, the model performs with different accuracy level for different subjects. Thus, fine-tuning is needed to ensure the reliability of the model.

% However, not only different types of cohorts, devices, or EEG electrode placements resulted in diverged collected signals, various subjects or recording sessions can also impact the variability \cite{buckelmuller2006trait}. Therefore, there is no safeguard for the efficiency of a model performance in every incoming subject despite the fine-tuning of some subjects from particular cohorts.

\section{Motivation and Contribution}
\label{sec:motivation}
Several works have paved a way to personalize, i.e., adjusting the model to serve for each subject. For example, instead of transferring knowledge from large to small datasets, Mikkelsen \textit{et al.} \cite{mikkelsen2018personalizing} proposed a personalized model and pre-trained 19 subjects. The pre-training was referred as a generalized model, following by the fine-tuning using the first night of one target subject within the same dataset. The model was evaluated on the individual’s data on the second night. Using more diverse data, Andreotti \textit{et al.} \cite{andreotti2018multichannel} transferred knowledge from a combination of two large datasets to other smaller datasets in both healthy subjects and patients. The model was personalized by fine-tuning each subject using 20 patients in the target cohort along with the first night of one target patient. The model was then tested on the second night data of that patient. The limitation of these two works, however, requires the training of the subjects with similar characteristic of sleep stages, i.e., subjects from the same cohort or having similar sleep-related disorders to the target subject. Another study recently published by Phan \textit{et al.} \cite{phan2020personalized}, involved the fine-tuning of the pre-trained SeqSleepNet, which was performed on a large dataset, to each subject in another small dataset by using only the first night of each target subject. KL-divergence regularization was applied in order to avoid the over-fitting issue due to the large network though small fine-tuning data. The results showed that personalizing improved the performance on sleep stage classification on each subject's data. However, the data recorded on patients with sleep disorders who have different sleep characteristics are usually more difficult to classify \cite{norman2000interobserver, korkalainen2019accurate} and have not been specifically observed. Furthermore, in many models, the first night of each target subject is still required for the performance.

In a different approach, Chen \textit{et al.} \cite{chen2019personalized} proposed the sleep stage personalizing based on Symbolic Fusion (SF) and Differential Evolution (DE). Firstly, a set of digital parameters based on the domain knowledge of sleep medicine, such as EEG sleep spindle, was extracted from the raw signals. The clinicians labelled sleep stages for only 5\% of each record. The DE method was applied to those samples to automatically generate the thresholds, which transformed those digital parameters to symbolic features (e.g., high, medium, and low). The symbolic features were then fed into the inference method in order to classify the sleep stages. Finally, the results were modified with their correction rules based on common patterns of sleep stages. The results were demonstrated to outperform the normal feature extraction methods with ML approach. Still, the classification of N1 stage yielded lower accuracy than the results from DL approach. Aside from the performance, there were also other limitations compared to DL approach: 1) Domain knowledge is required to frame the rules for feature extraction and sleep stage classification which can be done automatically by DL, 2) the performance does not rely on the classification algorithm only, but also the feature extraction method which might not be generalized as the team evaluated on their internal datasets including only 16 subjects, and 3) the decision rules were structured by domain knowledge, therefore, some remainders might not be included due to the variability of the ability to detect installed in the machine. In the learning system, not only the human can teach the machine, but the machine can detect some information insight which might not be definable by humans. Thus, the DL approach appears to be the best method in order to achieve high performance.

% Firstly, the features were extracted from raw signals. The clinicians labelled only 5\% of each record and used those samples to automatically generate the thresholds via DE, for transforming those features to symbolic features. The symbolic features were then fed into the inference method in order to classify the sleep stages. Finally, the results were modified with their correction rules based on common patterns of sleep stages. The results were shown to outperform the normal feature extraction methods with ML approach. Still, the classification of N1 stage yielded lower accuracy than the results from DL approach. Thus, the DL approach appears to be the best method in order to achieve higher quality performance.

In the present work, we presented a pilot study employing a framework, namely \textit{MetaSleepLearner}, to explore the feasibility of solving the aforementioned limitations. Firstly, we used the DL-based approach to develop and possibly expand to achieve the highest performance. We also aimed to solve the issue of limited numbers of data by using TL approach. Secondly, the differences in the recording of each subject were solved by transferring knowledge from pre-trained subjects to new individuals. Thirdly, to mitigate the time-consuming problem from manually labelling the signals recorded the whole night, we motivate and encourage the collaboration between clinicians and machines. The proposed framework allows clinicians to label only several samples per sleep stage, while the remaining labelling can be done automatically by the system. To accomplish our goal, an advanced TL method called Model Agnostic Meta-Learning (MAML) \cite{finn2017model}, were applied in our approach. To our best knowledge, MAML, which has been widely used in computer visions, has never been used in sleep stage classification. While all previous sleep stage TL works have been focusing on the performance in each state-of-the-art network or different fine-tuning paradigms, we instead attempted to elicit the performance of MAML in the pre-training phase. Its capability includes pre-training on various tasks and is claimed to be an algorithm with fast adaptation to the new tasks by using only a few samples. Therefore, the comparison between the conventional TL method and MAML were mainly investigated. Lastly, the results on both healthy subjects and patients were investigated, in order to ensure the performance on different variations of subjects.

\begin{figure*}
\centering
  \includegraphics[width=0.9\textwidth]{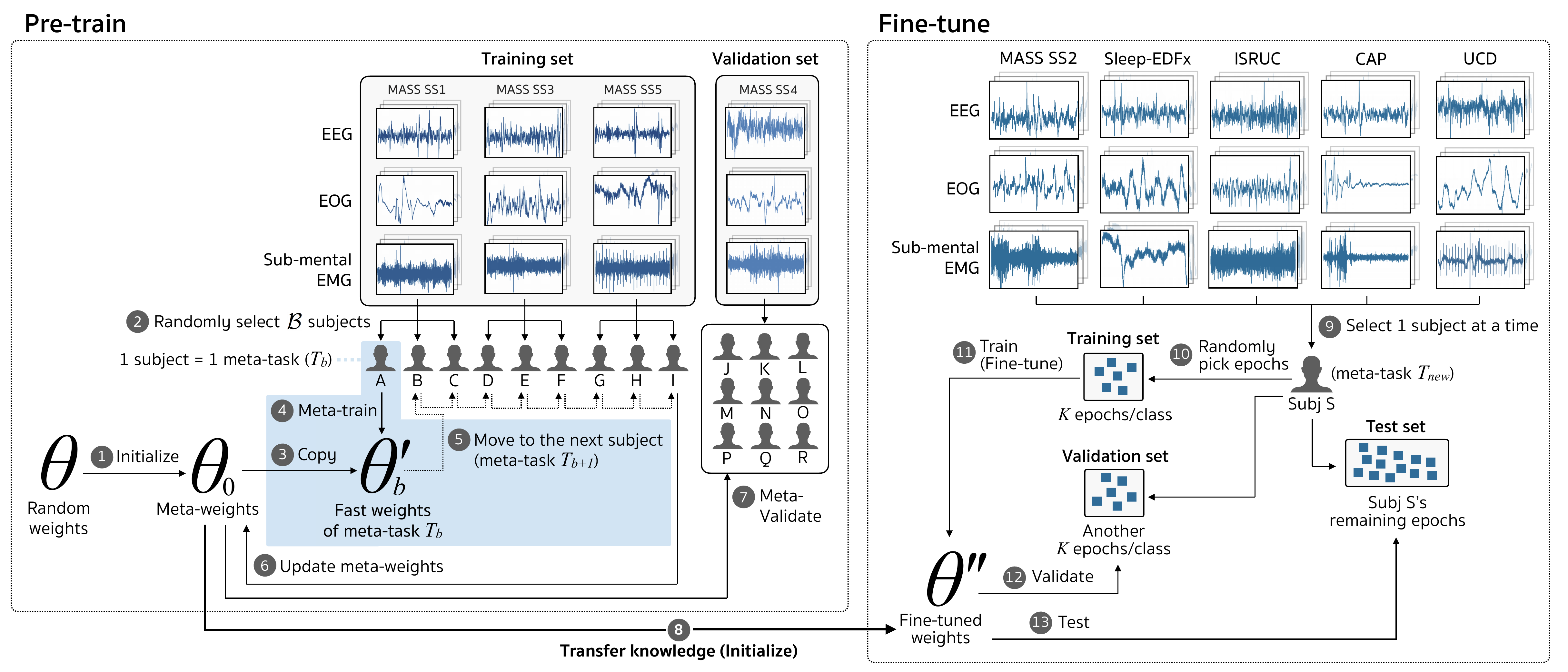}
  \caption{\textit{MetaSleepLearner}. Our proposed framework composes of two phases: Pre-training \& Fine-tuning. The process of pre-training begins by the initialization of the meta-weights, abbreviated as $\theta_0$, from randomized weights to the deep neural network (DNN) model. Then, the meta-training was performed using the data from MASS dataset resulting a pre-trained version of $\theta_0$. The pre-training phase ends when the model acceptably trains the $\theta_0$ to be fitted to the data extracted from the validation set. The second phase of fine-tuning commences to test the performance of $\theta_0$ by transferring the knowledge from the pre-training phase ($\theta_0$) to fine-tune a subject from other cohorts. The model ($\theta''$ initialized by $\theta_0$ from the pre-training phase) is then fine-tuned using the training set and tuned hyperparameters with the validation set. Ultimately, the performance of $\theta''$ is evaluated using the test set, concluding the ability of the fine-tuning phase to be used with any subject with altered conditions.}
  \label{fig:overview}
\end{figure*}

\section{Method}
\label{sec:method}

\subsection{Transfer Learning in Deep Learning}
\label{sec:TL}
Many DL researchers and practitioners do not have enough computational resources or sufficient data to train DL models from scratch. It has been found to be more practical to adapt existing models, which are shared in the community, to the task at hand. This practice, namely \emph{Transfer Learning (TL)}, describes heuristics of how to consolidate knowledge from one task (i.e., pre-training) and unpack it in other learning tasks (i.e., adaptation). It has become very influential in many fields, such as robotics \cite{arulkumaran2016classifying,james2017transferring}, computer visions \cite{redmon2016you,hu2018learning}, and natural language processing \cite{devlin2018bert,young2018recent,yang2019xlnet}. 

% However, the pre-training phase in the conventional TL paradigm tries to reach the global optimum from the given data. Such that, the adaptation on other data such as other subjects, which can be different in many aspects in real-word situation, is challenging.

% TL has been used with three major components of DL algorithm: (i) the pre-trained weights \cite{redmon2016you, hu2018learning}, (ii) the network architecture \cite{brock2018smash}, and (iii) the learning algorithm and optimization \cite{andrychowicz2016learning}. These components influence the learning by biasing the model in different ways. For example, neural network's weights can put the neural network model close to a certain area in the optimization landscape, encouraging the learning to converge at a certain local minima.

\subsection{Model Agnostic Meta-Learning}
In this work, we focused on an advanced TL method, called MAML \cite{finn2017model}. Normally, TL paradigms compose of 2 phases: pre-training and adaptation (or fine-tuning). MAML describes an algorithm that learns a set of pre-trained weights, which can be easily adapted to new tasks, in contrast to the conventional one, which tries to reach the global optimum from the given data. The benefits of MAML go beyond the reuse of features because, while consolidating, MAML considers the possible changes in these features in the adaptation phase. In other words, MAML grasps (in pre-training phase) at how to quickly learn or adapt in related tasks. We call this ability of learning to learn -- \emph{meta-learning}.

This meta-learning ability of MAML is interesting for our task because, unlike image data, the useful features of the bio-signals are less understood. While it is easy to reuse the primitives in image data, it is unclear whether a bio-signal feature from one person (or one device) can be reused, without adaptation, for another person (or another device). Moreover, the MAML itself claims to be a fast adaptation method, which serves our goal, i.e., requiring only a few samples from new individuals for adaptation.

Formally, a neural network, $f_{\theta}$, is parameterized by its parameters or model's weights, $\theta$. The objective of MAML is to find an initial $\theta = \theta_0$, called meta-weights, that, after updated with stochastic gradient descent ($\theta_b' \leftarrow \theta_0$), performs well on a set of tasks $\{T_0, T_1, .., T_B\}$). In other words, it finds $\theta_0$, which returns the sum of the objective values, after updated, as low as possible. Therefore, $\theta_0$ has to be easily adaptable, i.e., it has to be easily updated into a set of good $\theta_b'$.  

The main hypothesis assumes that $\theta_0$ will be able to generalize to unseen tasks $T_{\text{new}}$, which are from the same distribution as ${T_1, T_2, .., T_B}$. Therefore, by training $f_{\theta}$ on a set of related tasks, we consolidated the knowledge of these tasks into $\theta_0$, which could then be easily adapted (or fine-tuned) to an unseen task.

\begin{algorithm}[t]
% \onehalfspacing
\SetAlgoLined
\SetKwInOut{Require}{\textbf{Require}}
\SetKwFor{ForEach}{\textbf{for each}}{do}{\textbf{end for}}
\SetKwFor{While}{\textbf{while}}{do}{\textbf{end while}}
\SetKwFor{For}{\textbf{for}}{do}{\textbf{end for}}
\Require{Meta-training task $T_{train}$}{}
    \ForEach{meta-training iteration}{
        Sample training tasks $T_b \in T_{train}, 1 \leq b \leq \mathcal{B}$
        
        \ForEach{meta-task $T_b$}{
          Copy: $\theta_b' \gets \theta_0$\
          
          $\text{Randomly pick } K \text{ epochs / sleep stage from } T_b \text{ for 2 sets:} $
          
          $D_p = \{x_i, y_i\} \text{ and } D_q = \{x_j, y_j\} $
        
          \For{\texttt{$step = 1 \rightarrow num\_updates$}}{
            Predict using $D_p: \forall{i}, \text{ } f_{\theta'_b}(x_i)$
            
            Calculate loss: $\mathcal{L}_{T_b}(f_{\theta'_b}) $
            
            Update fast weights using \autoref{eq:fast_weight_update}: 
            $\theta_b' \leftarrow \theta_b' - \alpha \nabla _{\theta_b'}\mathcal{L}_{T_b}(f_{\theta'_b})  $
            
            Predict using $D_q: \forall{j}, \text{ }f_{\theta_b'}(x_j) $
            
            Calculate loss: $\mathcal{L}_{T_b}(f_{\theta'_b}) $
            
            Keep $\mathcal{L}_{T_b}$ for further calculation:
            $\mathbb{L}_{T_b}.append(\mathcal{L}_{T_b})$
          } 
    }
    Sum loss from all meta-tasks using \autoref{eq:loss_meta}:
    $\mathcal{L}_{\text{meta}}(\theta_0) = \sum_{b=1}^{B} \mathbb{L}_{T_b}(f_{\theta'_b})$
    
    Update meta-weights using \autoref{eq:update_theta0}:
    $\theta_0 \leftarrow \theta_0 - \gamma \nabla_{\theta_0} \mathcal{L}_\text{meta}(\theta_0)$
}

\caption{\hbox{Meta-training for Fast-Adaptation}}
\label{pseudocode}
\end{algorithm}

\subsection{MetaSleepLearner}
\label{sec:fast-adapt}

This pilot study aimed to explore the feasibility of employing MAML to perform fast adaptation of sleep staging to new individuals. The overall process, consisting of two phases, is illustrated in \autoref{fig:overview}. The input data included different channels of bipolar EEG, EOG, and submental EMG, which are described in Appendix \ref{app:data}. Due to the variability of bio-signals in each recording and each subject, mentioned in \autoref{sec:introduction}, the model was expected to learn and adapt to each of them, i.e., fine-tuned to each record of individual subjects. Therefore, one meta-task (referred as $T_b$) represents one record. Using a large standard sleep dataset (${T_1, T_2, .., T_B}$), the model pre-trained with our approach, called \textit{meta-train}, yielded $\theta_0$ as a result. The knowledge, consolidated on the set of weight $\theta_0$, were then transferred to $T_{new}$, which represented each new individual in other incoming datasets. The details of training steps are described as follows.

%This pilot study aimed to explore the feasibility of employing MAML to perform fast adaptation of sleep staging to new individuals. The overall process, consisting of two phases, is illustrated in \autoref{fig:overview}. The input data included different channels of bipolar EEG, EOG, and submental EMG, which are described in Appendix \ref{app:data}. \textcolor{red}{Due to the variability of bio-signals in each recording and each subject, mentioned in \autoref{sec:introduction}, the model should learn to adapt to each of them, i.e. fine-tuned to each record of individual subjects. Therefore, one meta-task (referred as $T_b$) represents one record.} The model pre-trained with our approach, \textcolor{red}{called} \textit{meta-train}, on a large standard sleep dataset (${T_1, T_2, .., T_B}$), yielded $\theta_0$ as a result. The knowledge, consolidated on the set of weight $\theta_0$, were then transferred to $T_{new}$, which represented each new individual in other incoming datasets. The details of training steps are described as follows.

\subsubsection{Pre-train (Meta-training)}

We randomly selected 3 subsets of MASS \cite{mass} for pre-training (SS1, SS3, and SS5) and another one (SS4) for validation. Each subject in each subset was treated as one \textit{meta-task} ($T_b$), in which the tasks were divided into two groups: for training ($T_{train}$) and for validation ($T_{val}$). The model weights, referred to as meta-weights ($\theta_0$), were firstly initialized randomly using Xavier \cite{xavier2010}, a randomized weight initialization method. 

As described in \autoref{pseudocode} and illustrated in \autoref{fig:meta-train} (Appendix \ref{app:meta-train}), in each meta-training iteration, $\mathcal{B}$ tasks were randomly selected for meta-training (${T_1, T_2, .., T_B \in T_{train}}$). In order to adapt to those selected tasks, each $T_b$ copied the weights from $\theta_0$ as its own fast-weights ($\theta_b'$):

\begin{equation}
 \theta'_b \leftarrow \theta_0,
 \label{eq:copy_weight}
\end{equation}

Subsequently, two sets of samples were randomly selected from $T_b$, referred to as $D_p$ and $D_q$, each of which consisted of $K$ epochs per sleep stage, i.e., the variable number of samples per sleep stage. Thus, for five sleep stages, a total of $K \times 5$ epochs were chosen per set. In order to adapt to those tasks, the gradient descent was performed separately for $num\_updates$ steps. Each step starting from using $D_p$ to update the $\theta_b'$ was shown as follows:

\begin{equation}
 \theta'_b \leftarrow \theta'_b - \alpha \nabla_{\theta'_b} \mathcal{L}_{T_b}(f_{\theta'_b})
 \label{eq:fast_weight_update}
\end{equation}

where, $\alpha$ is an updating step size or learning rate ($update\_lr$). The updated $\theta'_b$ was evaluated using $D_q$, referred as $\mathcal{L}_{T_b}$, which would eventually be kept in a list $\mathbb{L}_{T_b}$. After all $\mathcal{B}$ selected meta-tasks were executed, the summation of losses were calculated as:

\begin{align} 
% \mathcal{\min_{\theta_0}} \text{ }
\mathcal{L}_{\text{meta}}(\theta_0) = \sum_{b=1}^{\mathcal{B}} \mathbb{L}_{T_b}(f_{\theta'_b}),
\label{eq:loss_meta}
\end{align}

which is the objective function of MAML. As a result of a succession of meta-training iteration, $\theta_0$ was updated with the gradient descent of $\mathcal{L}_{\text{meta}}$ using

\begin{equation}
  \theta_0 \leftarrow \theta_0 - \gamma \nabla_{\theta_0} \mathcal{L}_\text{meta}(\theta_0),
  \label{eq:update_theta0}
\end{equation}

where $\gamma$ is the learning rate of updating meta-weights ($meta\_lr$). Following the re-calculation of $\theta_0$, the meta-validation was performed by sampling the tasks from $T_{val}$. The sequence procedures abided by the process of meta-training %Noted that the $\mathcal{L}_{T_b}$ from meta-validation was not denoted for updating $\theta_0$,%
and only for tuning the hyperparameters, e.g. number of meta-training iterations, learning rates, etc. This meta-validation procedure benefits the pre-training process for fast adaptation because the $\theta_0$ is not selected when it immediately performs well with the validation set. Instead, it is selected when it achieves good validation loss after adapting to those data, resulting in the effective adaptation.

\subsubsection{Fine-tune (Adaptation)} This phase is considered the conventional adaptation method in TL. The knowledge from pre-training phase was transferred to this phase by initializing the fine-tuned weights ($\theta''$) with $\theta_0$. The objective of this study is to quickly adapt to new individuals in the new cohorts. Hence, one-night record from one subject was selected at a time from the unseen cohorts, referred as $T_{\text{new}}$. The sleep epochs, i.e. samples of 30-second long of raw signals, extracted from the selected subject, were randomly divided into three different sets: a set of \textit{K} epochs per sleep stage as a training set, another set of \textit{K} epochs per sleep stage for validation, and the remaining epochs for testing. The adaptation procedure was similar to the adaptation in each meta-training iteration. The gradient descent was performed to adapt the $\theta''$ to each individual subject, using only \textit{K} epochs per sleep stage in training set, i.e., only a small amount of sleep stage labels are required. The $\theta''$ was then validated against the other \textit{K} epochs per sleep stage from the validation set in order to find the suitable hyperparameters. Ultimately, the performance of the network was examined using the remaining epochs.

\section{Experiments}
\label{sec:exp}
To explore the feasibility of employing MAML in this task, we performed the experiments to support our hypothesis as follows. Firstly, we hypothesized that using TL, i.e., the pre-training of the model from large dataset and the fine-tuning on each target subject, should achieve higher performance than training the model with the target subject's data from scratch. Secondly, using multi-modals (EEG, EOG, and sub-mental EMG) in TL would be more advantageous than only EEG signals for sleep stage classification. Thirdly, the main experiment, the efficiency of our pre-training approach should be more efficiently than the conventional approaches in various sleep cohorts of both healthy subjects and patients with sleep disorders. Lastly, the knowledge acquired from our approach should be reasonable and explainable. In all experiments, the procedure was divided into 2 steps: pre-training and fine-tuning. The experimental setups are described in this section.

% To explore the feasibility of employing MAML in this task, we performed the experiments, aiming to achieve several goals as follows. Firstly, we showed the advantage of TL compared to training the model from scratch. Secondly, the advantages of using multi-modals (EEG, EOG, and sub-mental EMG) in TL instead of only EEG signals for sleep stage classification were explored. Thirdly, the efficiency of our pre-training approach was compared against the conventional approaches by using various sleep cohorts of both healthy subjects and patients with sleep disorders. Lastly, the model interpretation was illustrated to investigate the knowledge the model acquired from our approach. In all experiments, the procedure was divided into 2 steps: pre-training and fine-tuning. The experimental setups are described in this section.

\subsection{Datasets}
Five publicly available datasets were used in our experiments: 1) MASS \cite{mass} was published as a common benchmark dataset for sleep research. The performance took place at three different hospital-based sleep laboratories in Canada, recorded from 200 participants, and separated into five subsets. 2) Sleep-EDF \cite{kemp2000analysis} from MCH-Westeinde Hospital, Den Haag, Netherlands, consists of two subsets; SC is a study of age effects on sleep in 20 healthy subjects and ST is a study of the effects of the drug temazepam during sleep in 22 patients with mild falling asleep difficulty. 3) The creator of CAP sleep database \cite{terzano2001atlas} aimed to investigate the relation between Cyclic Alternating Pattern (CAP) and pathologies in 108 polysomnographic recordings registered at the Sleep Disorders Center of the Ospedale Maggiore of Parma, Italy. 4) ISRUC \cite{isruc} contains both healthy subjects and subjects with sleep disorders at the Sleep Medicine Centre of the Hospital of Coimbra University (CHUC), which was created for sleep researchers. 5) UCD encompasses 25 subjects with suspected sleep-disordered breathing problem at the Sleep Disorders Clinic, St Vincent's University Hospital, Dublin \cite{goldberger2000physiobank}. Further descriptions of each dataset are provided in Appendix \ref{app:data}.

\subsection{Model Specification}
The goals of the experiments were not to compare with other network architectures, but mainly to test the hypothesis that our approach (meta-training) would yield better results than the conventional pre-training methods. Therefore, the experiments were executed using an identical network architecture on each training paradigm for a direct assessment. In order to reduce the computational time and resources, we performed the experiments using a simplified version of CNN network, based on the state-of-the-art model, namely DeepSleepNet \cite{supratak2017deepsleepnet}, as described in Appendix \ref{app:model}.

\subsection{Pre-training}
The experimental setups of the first step of TL, i.e., pre-training, are described in this section. As mentioned earlier in \autoref{sec:fast-adapt}, for each meta-task $T_b$, i.e., each subject, $K$ epochs per sleep stage were randomly chosen for training and other $K$ epochs per sleep stage for validation. The subjects, whose sleep stages were less than $K \times 2$ epochs, were filtered out. To satisfy our goal which allows the clinicians to label only a few epochs of a PSG recording, reducing the strain, the $K$ was set as 10. After filtering the number of epochs per sleep stage, there were 30, 37, and 24 subjects from MASS SS1, SS3, and SS5, respectively, as parts of the training set. For particular subsets which segmented each epoch for 20 seconds long, we appended 5-second segments of its preceding and succeeding epochs to make a 30-second epoch. More details of data pre-processing are described in Appendix \ref{app:data}. MASS was selected for the pre-training due to its large amount of data compared to the other public datasets. In basic PSG and all acquired cohorts, C3 is one of the most commonly used EEG electrode placements, in accordance with the International 10–20 system. Hence, the bipolar C3-A2 EEG electrodes were chosen as samples along with left - right EOG and sub-mental EMG provided by each subset. To compare the performance of our approach against all baselines, the pre-training details of each approach were set as follows:
% In order to examine the performance of our proposed method, the pre-training phase was executed in an identical network architecture with all pre-training paradigms.

\subsubsection{Our proposed Approach (MetaSleepLearner)} 
The meta-training procedures are described in \autoref{sec:fast-adapt}. The number of meta-tasks selected in each meta-iteration ($\mathcal{B}$) was set to 9. For meta-validation, to be comparable with the number of meta-tasks, 9 subjects with sufficient number of recorded epochs per sleep stage were randomly selected from MASS SS4 and fixed as the representatives in every model run. The model was meta-trained until the increasing trend of meta-validation loss was observed and the model's weights (meta-weights $\theta_0$) were kept at the best iteration. The set of hyperparameters included the learning rate at the updating meta-weights ($meta\_lr$ ($\gamma$) $\in$ \{$10^{-2}, 10^{-3}, 10^{-4}$\}), the learning rate inside the sub-task adaptation ($update\_lr$ ($\alpha$) $\in$ \{$10^{-2}, 10^{-3}, 10^{-4}$\}), and the number of updating steps ($num\_updates$ $\in$ \{5, 10, 15\}).

\subsubsection{Conventional Approach (Baseline-1)} 
The conventional pre-training method was used as a baseline in comparison to our approach. The training procedure involved all samples in the training datasets as they were pooled towards one place. To avoid the class-imbalanced issue, the training data were over-sampled by duplicating under-present sleep stages to achieve an equal number of epochs per sleep stage, in which the validation loss was found to be lower than the non-oversampling one. In each training iteration, the model randomly drew upon the samples to train using mini batches, performing the gradient descent in the same network architecture as our approach. After all mini batches were put through the network, one training iteration ended. The validation was then performed by utilizing all samples from the validation set, which were from the same set of subjects as meta-validation. The model was trained repeatedly until the validation loss did not improve for at least 100 iterations. The best iteration was maintained as $\theta_0$. The sets of hyperparameters included the $learning\_rate$ $\in$ \{$10^{-1}, 10^{-2}, 10^{-3}, 10^{-4}, 10^{-5}$\}, and $batch\_size$ $\in$ \{256, 512, 1024, 2048\}.

\subsubsection{Conventional Approach with one-batch training (Baseline-2)}
In Baseline-1, since the large amount of pre-training datasets was put into the network, the conventional training approach tended to fit at very early iterations. Moreover, we also altered to select only the same number of training and validating samples in each training iteration in order to compare directly to our approach. The validation samples were also extracted from the same set of meta-validation subjects. However, instead of using all samples from the validation set as in Baseline-1, we randomly selected only $K \times 2$ or 20 epochs per sleep stage from each subject as our approach. The model was also trained repeatedly until the validation loss did not improve for at least 200 iterations and kept the best iteration as $\theta_0$. The sets of hyperparameters were the same as in Baseline-1, with the exception of $batch\_size$, which were from \{250, 350, 450\}. It was deemed logical to use this set of $batch\_size$ in order to equalize it to our proposed approach by arranging $K \times 9$ per sleep stage = 450. Additionally, the number multiplying to $K$ was also varied ($K \times 5$ = 250 and $K \times 7$ = 350).

In each pre-training paradigm, all combinations of hyperparameters were performed. The set achieving the lowest validation loss was selected subsequently. The model was performed 5 times with the selected hyperparameters, giving rise to 5 sets of $\theta_0$ from MetaSleepLearner, 5 sets of $\theta_0$ from Baseline-1, and 5 sets of $\theta_0$ from Baseline-2.

\subsection{Fine-Tuning}
The knowledge from the first phase were then transferred to the fine-tuning phase for adaptation to new individuals in new cohorts. To fine-tune the model, we used four types of weights initialization for comparison: 1) random initialization using Xavier (training from scratch), 2) $\theta_0$ from MetaSleepLearner, 3) $\theta_0$ from Baseline-1, and 4) $\theta_0$ from Baseline-2. The same fine-tuning procedures were applied to all weights initializations in order to compare the quality, i.e., the adaptability, of pre-trained weights from each approach.

Each model was fine-tuned to each individual subject from the unseen datasets composing of both healthy subjects (as reported in their datasets) and patients, including Sleep-EDF, CAP, ISRUC, UCD, and the remaining subset from MASS (SS2). The EEG electrode placements were used differently, as shown in \autoref{tab:results}. The main channel was C3-A2. However, due to some records whose C3-A2 channels were not available from the dataset, we used C3-P3 or C4-A1 instead because their characteristics are comparable.

In the fine-tuning phase, the validation set was used to select the most suitable $learning\_rate \in \{10^{-1}, 10^{-2}, 10^{-3}, 10^{-4},  10^{-5}\}$, as well as the number of training iterations. The default $K$ and maximum training iterations were set to 5 and 500, respectively. However, in some patient datasets, the performances after evaluating every approach were very poor. Thus, we extended the $K$ to 10 and the maximum number of training iterations to 1000. In order to demonstrate that the model could be applied to an actual arbitrary situation, e.g., the random selection of any epochs labelled by the clinicians, the model randomly selected samples to run for 5 times per indicated subject and per weight initialization. For the comparison between the performances of all pre-training paradigms, the samples, which were randomly picked for training, validating, and testing during the same round from the same subject, were identical.

\begin{table*}[]
\centering
\fontsize{19}{24}\selectfont 
\caption{Performance of five sleep stages classification after fine-tuning on each individual in each cohort. The results were averaged from all 5 pre-trained weights per each paradigm $\times$ no. of subjects $\times$ 5 times of random samples selection. The subjects reported in this table are only those who contained enough epochs for fine-tuning, validating, and testing in their PSG recording.}
\label{tab:results}
\resizebox{\textwidth}{!}{%
\begin{threeparttable}
\begin{tabular}{|l|c|c|ccc|ccc|ccc|ccc|ccc|ccc|ccc|ccc|}
\hline
\multicolumn{1}{|c|}{\multirow{3}{*}{\textbf{Dataset}}} & \multirow{3}{*}{\textbf{\begin{tabular}[c]{@{}c@{}}EEG\\ channel\end{tabular}}} & \multirow{3}{*}{\textbf{\begin{tabular}[c]{@{}c@{}}No. of\\ subjects\end{tabular}}} & \multicolumn{3}{c|}{\multirow{2}{*}{\textbf{\begin{tabular}[c]{@{}c@{}}Overall \\ Accuracy (\%)\end{tabular}}}} & \multicolumn{15}{c|}{\textbf{F1 per class}} & \multicolumn{3}{c|}{\multirow{2}{*}{\textbf{MF1}}} & \multicolumn{3}{c|}{\multirow{2}{*}{\textbf{Cohen's kappa ($\mathcal{K}$)}}} \\ \cline{7-21}
\multicolumn{1}{|c|}{} &  &  & \multicolumn{3}{c|}{} & \multicolumn{3}{c|}{\textbf{W}} & \multicolumn{3}{c|}{\textbf{N1}} & \multicolumn{3}{c|}{\textbf{N2}} & \multicolumn{3}{c|}{\textbf{N3}} & \multicolumn{3}{c|}{\textbf{REM}} & \multicolumn{3}{c|}{} & \multicolumn{3}{c|}{} \\
\multicolumn{1}{|c|}{} &  &  & \textbf{B-1} & \textbf{B-2} & \textbf{ours} & \textbf{B-1} & \textbf{B-2} & \textbf{ours} & \textbf{B-1} & \textbf{B-2} & \textbf{ours} & \textbf{B-1} & \textbf{B-2} & \textbf{ours} & \textbf{B-1} & \textbf{B-2} & \textbf{ours} & \textbf{B-1} & \textbf{B-2} & \textbf{ours} & \textbf{B-1} & \textbf{B-2} & \textbf{ours} & \textbf{B-1} & \textbf{B-2} & \textbf{ours} \\ \hline \hline
\textbf{Healthy} & \multicolumn{1}{l|}{} & \multicolumn{1}{l|}{} & \multicolumn{1}{l}{} & \multicolumn{1}{l}{} & \multicolumn{1}{l|}{} & \multicolumn{1}{l}{} & \multicolumn{1}{l}{} & \multicolumn{1}{l|}{} & \multicolumn{1}{l}{} & \multicolumn{1}{l}{} & \multicolumn{1}{l|}{} & \multicolumn{1}{l}{} & \multicolumn{1}{l}{} & \multicolumn{1}{l|}{} & \multicolumn{1}{l}{} & \multicolumn{1}{l}{} & \multicolumn{1}{l|}{} & \multicolumn{1}{l}{} & \multicolumn{1}{l}{} & \multicolumn{1}{l|}{} & \multicolumn{1}{l}{} & \multicolumn{1}{l}{} & \multicolumn{1}{l|}{} & \multicolumn{1}{l}{} & \multicolumn{1}{l}{} & \multicolumn{1}{l|}{} \\ \hline
\textbf{1D-CNN} & \multicolumn{1}{l|}{} & \multicolumn{1}{l|}{} & \multicolumn{1}{l}{} & \multicolumn{1}{l}{} & \multicolumn{1}{l|}{} & \multicolumn{1}{l}{} & \multicolumn{1}{l}{} & \multicolumn{1}{l|}{} & \multicolumn{1}{l}{} & \multicolumn{1}{l}{} & \multicolumn{1}{l|}{} & \multicolumn{1}{l}{} & \multicolumn{1}{l}{} & \multicolumn{1}{l|}{} & \multicolumn{1}{l}{} & \multicolumn{1}{l}{} & \multicolumn{1}{l|}{} & \multicolumn{1}{l}{} & \multicolumn{1}{l}{} & \multicolumn{1}{l|}{} & \multicolumn{1}{l}{} & \multicolumn{1}{l}{} & \multicolumn{1}{l|}{} & \multicolumn{1}{l}{} & \multicolumn{1}{l}{} & \multicolumn{1}{l|}{} \\
\textbf{EEG only} & \multicolumn{1}{l|}{} & \multicolumn{1}{l|}{} & \multicolumn{1}{l}{} & \multicolumn{1}{l}{} & \multicolumn{1}{l|}{} & \multicolumn{1}{l}{} & \multicolumn{1}{l}{} & \multicolumn{1}{l|}{} & \multicolumn{1}{l}{} & \multicolumn{1}{l}{} & \multicolumn{1}{l|}{} & \multicolumn{1}{l}{} & \multicolumn{1}{l}{} & \multicolumn{1}{l|}{} & \multicolumn{1}{l}{} & \multicolumn{1}{l}{} & \multicolumn{1}{l|}{} & \multicolumn{1}{l}{} & \multicolumn{1}{l}{} & \multicolumn{1}{l|}{} & \multicolumn{1}{l}{} & \multicolumn{1}{l}{} & \multicolumn{1}{l|}{} & \multicolumn{1}{l}{} & \multicolumn{1}{l}{} & \multicolumn{1}{l|}{} \\
Sleep-EDF (SC) & Fpz-Cz & 18 & 59.0 & 60.5 & \textbf{61.4} & 53.2 & 57.7 & 58.5 & 16.1 & 20.2 & 19.6 & 64.9 & 67.3 & 67.7 & 65.8 & 67.8 & \textbf{70.7} & 52.6 & 54.3 & 54.8 & 50.5 & 53.4 & \textbf{54.3} & 0.446 & 0.472 & \textbf{0.482} \\
\textbf{2D-CNN} & \multicolumn{1}{l|}{} & \multicolumn{1}{l|}{} & \multicolumn{1}{l}{} & \multicolumn{1}{l}{} & \multicolumn{1}{l|}{} & \multicolumn{1}{l}{} & \multicolumn{1}{l}{} & \multicolumn{1}{l|}{} & \multicolumn{1}{l}{} & \multicolumn{1}{l}{} & \multicolumn{1}{l|}{} & \multicolumn{1}{l}{} & \multicolumn{1}{l}{} & \multicolumn{1}{l|}{} & \multicolumn{1}{l}{} & \multicolumn{1}{l}{} & \multicolumn{1}{l|}{} & \multicolumn{1}{l}{} & \multicolumn{1}{l}{} & \multicolumn{1}{l|}{} & \multicolumn{1}{l}{} & \multicolumn{1}{l}{} & \multicolumn{1}{l|}{} & \multicolumn{1}{l}{} & \multicolumn{1}{l}{} & \multicolumn{1}{l|}{} \\
\textbf{EEG only} &  &  &  &  &  &  &  &  &  &  &  &  &  &  &  &  &  &  &  &  &  &  &  &  &  &  \\
Sleep-EDF (SC) & Fpz-Cz & 18 & 64.9 & \textbf{73.6} & 72.1 & 59.6 & \textbf{75.2} & 70.0 & 21.1 & 28.7 & \textbf{29.0} & 70.5 & \textbf{79.2} & 78.6 & 74.3 & 79.2 & 79.1 & 60.6 & \textbf{69.7} & 67.5 & 57.2 & \textbf{66.4} & 64.8 & 0.536 & \textbf{0.644} & 0.624 \\
\textbf{EEG, EOG, Submental EMG} &  & \multicolumn{1}{l|}{} & \multicolumn{1}{l}{} & \multicolumn{1}{l}{} & \multicolumn{1}{l|}{} & \multicolumn{1}{l}{} & \multicolumn{1}{l}{} & \multicolumn{1}{l|}{} & \multicolumn{1}{l}{} & \multicolumn{1}{l}{} & \multicolumn{1}{l|}{} & \multicolumn{1}{l}{} & \multicolumn{1}{l}{} & \multicolumn{1}{l|}{} & \multicolumn{1}{l}{} & \multicolumn{1}{l}{} & \multicolumn{1}{l|}{} & \multicolumn{1}{l}{} & \multicolumn{1}{l}{} & \multicolumn{1}{l|}{} & \multicolumn{1}{l}{} & \multicolumn{1}{l}{} & \multicolumn{1}{l|}{} & \multicolumn{1}{l}{} & \multicolumn{1}{l}{} & \multicolumn{1}{l|}{} \\
Sleep-EDF (SC) & Fpz-Cz & 18 & 70.7 & 70.8 & \textbf{74.9} & 72.2 & 72.9 & \textbf{77.1} & 30.3 & 31.6 & \textbf{35.9} & 76.1 & 75.6 & \textbf{79.4} & 77.7 & 75.9 & \textbf{80.2} & 64.1 & 67.1 & \textbf{71.5} & 64.1 & 64.6 & \textbf{68.8} & 0.606 & 0.610 & \textbf{0.662} \\
ISRUC (Subgroup 3) & C3-A2 & 10 & 72.3 & 70.7 & \textbf{75.2} & 73.1 & 71.2 & \textbf{78.9} & 39.6 & 39.2 & \textbf{43.6} & 71.6 & 69.4 & \textbf{73.5} & 83.2 & 81.1 & 83.6 & 71.1 & 70.5 & \textbf{74.9} & 67.7 & 66.3 & \textbf{70.9} & 0.633 & 0.613 & \textbf{0.671} \\
MASS (SS2) & \begin{tabular}[t]{@{}c@{}}C3-A2 /\\ C4-A1\end{tabular} & 19 & 74.7 & 72.4 & \textbf{77.3} & 68.2 & 66.7 & \textbf{73.7} & 28.0 & 28.0 & \textbf{32.9} & 79.9 & 77.4 & \textbf{81.6} & 83.7 & 80.7 & \textbf{84.3} & 73.0 & 71.4 & \textbf{77.1} & 66.6 & 64.8 & \textbf{69.9} & 0.648 & 0.618 & \textbf{0.682} \\
CAP & \begin{tabular}[t]{@{}c@{}}C3-A2 /\\ C3-P3 /\\ C4-A1\end{tabular} & 6 & 71.3 & 70.0 & \textbf{75.1} & 69.6 & 68.2 & \textbf{75.2} & 22.5 & 22.5 & \textbf{27.2} & 75.8 & 74.2 & \textbf{78.7} & 81.1 & 78.5 & \textbf{82.6} & 68.5 & 68.7 & \textbf{74.4} & 63.5 & 62.4 & \textbf{67.6} & 0.611 & 0.596 & \textbf{0.661} \\ \hline \hline
\textbf{Patients} &  &  &  &  &  &  &  &  &  &  &  &  &  &  &  &  &  &  &  &  &  &  &  &  &  &  \\ \hline
\textbf{2D-CNN} &  & \multicolumn{1}{l|}{} & \multicolumn{1}{l}{} & \multicolumn{1}{l}{} & \multicolumn{1}{l|}{} & \multicolumn{1}{l}{} & \multicolumn{1}{l}{} & \multicolumn{1}{l|}{} & \multicolumn{1}{l}{} & \multicolumn{1}{l}{} & \multicolumn{1}{l|}{} & \multicolumn{1}{l}{} & \multicolumn{1}{l}{} & \multicolumn{1}{l|}{} & \multicolumn{1}{l}{} & \multicolumn{1}{l}{} & \multicolumn{1}{l|}{} & \multicolumn{1}{l}{} & \multicolumn{1}{l}{} & \multicolumn{1}{l|}{} & \multicolumn{1}{l}{} & \multicolumn{1}{l}{} & \multicolumn{1}{l|}{} & \multicolumn{1}{l}{} & \multicolumn{1}{l}{} & \multicolumn{1}{l|}{} \\
\textbf{EEG, EOG, Submental EMG} & \multicolumn{1}{l|}{} & \multicolumn{1}{l|}{} & \multicolumn{1}{l}{} & \multicolumn{1}{l}{} & \multicolumn{1}{l|}{} & \multicolumn{1}{l}{} & \multicolumn{1}{l}{} & \multicolumn{1}{l|}{} & \multicolumn{1}{l}{} & \multicolumn{1}{l}{} & \multicolumn{1}{l|}{} & \multicolumn{1}{l}{} & \multicolumn{1}{l}{} & \multicolumn{1}{l|}{} & \multicolumn{1}{l}{} & \multicolumn{1}{l}{} & \multicolumn{1}{l|}{} & \multicolumn{1}{l}{} & \multicolumn{1}{l}{} & \multicolumn{1}{l|}{} & \multicolumn{1}{l}{} & \multicolumn{1}{l}{} & \multicolumn{1}{l|}{} & \multicolumn{1}{l}{} & \multicolumn{1}{l}{} & \multicolumn{1}{l|}{} \\
Sleep-EDF (ST) & Fpz-Cz & 15 & 60.7 & 60.1 & \textbf{67.1} & 54.3 & 56.6 & \textbf{61.0} & 25.4 & 27.6 & \textbf{33.4} & 66.6 & 64.5 & \textbf{72.8} & 74.3 & 71.7 & \textbf{78.8} & 50.3 & 52.5 & \textbf{57.8} & 54.2 & 54.6 & \textbf{60.8} & 0.476 & 0.471 & \textbf{0.554} \\
ISRUC (Subgroup 1) & \begin{tabular}[t]{@{}c@{}}C3-A2 /\\ C3-M2\end{tabular} & 96 & 67.7 & 65.3 & \textbf{71.0} & 70.7 & 68.6 & \textbf{77.2} & 40.7 & 39.6 & \textbf{44.2} & 65.8 & 62.4 & \textbf{68.0} & 79.1 & 76.1 & \textbf{80.2} & 65.2 & 63.6 & \textbf{69.7} & 64.3 & 62.1 & \textbf{67.8} & 0.577 & 0.547 & \textbf{0.618} \\
CAP (Patients) & \begin{tabular}[t]{@{}c@{}}C3-A2 /\\ C3-P3 /\\ C4-A1\end{tabular} & 49 & 67.9 & 67.8 & \textbf{71.4} & 68.9 & 69.4 & \textbf{74.8} & 29.6 & 30.3 & \textbf{33.6} & 67.1 & 66.4 & \textbf{69.8} & 79.1 & 78.6 & \textbf{80.1} & 64.3 & 65.4 & \textbf{70.0} & 61.8 & 62.0 & \textbf{65.7} & 0.570 & 0.570 & \textbf{0.615} \\
UCD & C3-A2 & 22 & 49.1 & 50.5 & \textbf{56.3} & 44.1 & 46.6 & \textbf{51.1} & 24.6 & 25.9 & \textbf{27.5} & 48.5 & 49.9 & \textbf{60.4} & 61.7 & 64.7 & \textbf{69.0} & 34.1 & 37.8 & \textbf{42.7} & 42.6 & 45.0 & \textbf{50.1} & 0.345 & 0.364 & \textbf{0.429} \\ \hline
\end{tabular}%

% \caption*{\footnotesize The \textbf{bold} numbers represent the highest performance among all paradigms with significant difference from others (p $<$ 0.05). \\ B-1 = Baseline-1, B-2 = Baseline-2, ours = our proposed approach \\ 3 modals = EEG, EOG, and submental EMG}
\begin{tablenotes}[para,flushleft]
The \textbf{bold} numbers represent the highest performance among all paradigms with significant difference from others (p $<$ 0.05). \\ B-1 = Baseline-1, B-2 = Baseline-2, ours = our proposed approach
  \end{tablenotes}
  
\end{threeparttable}
}
\end{table*}

\subsection{Model Interpretation using Layer-wise Relevance Propagation (LRP)}

By applying the proposed fast adaptation procedure, the pre-trained model was adapted to the new incoming subject, i.e., $T_{new}$, in the fine-tuning phase. The fine-tuning and the validation were performed using only 5-10 epochs per sleep stage. Regardless of the performance, it is necessary to assess whether the method of learning is reasonable. Hence, the application of the Layer-wise Relevance Propagation (LRP) was used for examination. LRP describes a model interpretation method, allowing the reasons for making each decision to be seen, i.e., prediction. It was employed in the previous work to interpret the classification model using the EEG signals \cite{lrp-eeg}. In this study, LRP was employed, revealing the conception of the reasons that model used for each prediction. Initializing from the fine-tuned model ($\theta''$), LRP propagates backward from the output ($f_{\theta''}$) throughout all the layers, reaching the input layer. The $Relevance$ scores ($R$) was returned, which was found to be devised from the same shape as the input of the original model $(3000, 1, 3)$. Each value of $R$ indicates the reason behind the contribution of each sampling point from the signals to the decision of the model. The $R$ scores was computed as:

\begin{equation}
    R_i = \sum_j \frac{a_i w_{ij}}{\sum_{0,i} a_i w_{ij}} R_j,
\end{equation}

where $R_i$ is a $R$ of neuron $i$, $i$ and $j$ are two neurons of any consecutive layers, $a_i$ is an activation of neuron $i$, and $w_{ij}$ is the trained weight (parameter) connecting between neurons $i$ and $j$.

\section{Results}
The results elaborated in this section are divided into two parts. Subsection A describes the suitable hyperparameters and the achieved loss values from both the conventional and our proposed pre-training procedures. In Subsection B to E, the averaged evaluation of the performance after fine-tuning (adapting) to new individuals in the new cohorts are reported, as shown in \autoref{tab:results}. The number of evaluation results from each cohort, before being averaged, were 5 pre-trained weights $\times$ the number of subjects $\times$ 5 random times, e.g., 450 result samples for SC or 2400 samples for ISRUC (subgroup I). The performance of each experiment, mentioned in this section, was reported as Cohen's Kappa $\pm$ standard errors ($\mathcal{K}$ $\pm$ SE). The General Linear Model with Repeated Measures ($\alpha$ = 0.05, Confidence Interval = 95\%) was used to illustrate the significant differences between the results from each approach. It was found that when the number of test epochs per class was very low, the results were not quite suitable to interpret as one incorrectly predicted sample could highly affect the per-class performance metrics. Therefore, the results in \autoref{tab:results} which are mentioned in this section were from the subjects whose samples were sufficiently obtained for fine-tuning, validating, and testing (at least $K \times 3$ epochs per sleep stage). For those subjects whose number of epochs is less than the number mentioned above, the performance is separately reported in \autoref{tab:results2} in Appendix.

\subsection{Pre-training}
\label{ResultA}
In this Subsection, we investigated the results of pre-training phase from all procedures. In Baseline-1, $learning\_rate = 10^{-2}$ with $batch\_size$ = 1024 yielded the best set of hyperparameters, giving the lowest validation loss. From all 5 runs, the validation loss achieved a range of 0.65 to 0.85. Whereas in Baseline-2, $learning\_rate = 10^{-2}$ with $batch\_size$ = 450 were the best set and the achieved validation loss ranged from 0.66 to 0.75. In comparison, using our approach, the best validation loss achieved from $meta\_lr = 10^{-3}$, $update\_lr = 10^{-2}$, and $num\_updates = 10$. However, the results were not as sensitive to these exact values of the learning rates. The best validation loss was 0.46 and the highest was 0.52. The meta-validation procedure, defining whether the weights are suitable for adapting, instigated lower validation loss in comparison to using the conventional methods, when the loss was achieved once the adaptation reached 10 steps.

The average numbers of training iterations, each of 5 runs, from Baseline-1, Baseline-2, and our proposed approach were 8, 1320, and 10162. The averaged computational times were approximately 22 minutes, 6 hours 19 minutes, and 1 day 19 hours 24 minutes, respectively. It is imperative to note that Baseline-1 yielded the lowest training iteration. However, within each iteration, each mini batch contained several sub-iterations until the whole samples were expended. It was deemed usual as our approach required larger amount of iterations due to the fluctuating loss from the meta-training procedure. Since the model’s weights were updated separately from the 9 meta-tasks before updating the meta-weights in every iteration, the direction of the gradients was more variable. However, both loss values and number of pre-training iterations did not affect the practical usage as the approach could utilize the $\theta_0$ from this phase to adapt to the data from the new incoming cohorts in the fine-tuning phase.

\subsection{Advantages of Transfer Learning}
\label{ResultB}
Prior to inspecting the improvement of the proposed pre-training procedures, we examined whether the network trained using TL performed better than the neural network training from scratch. In this experiment, only EEG signals were used. CNN was then changed to 1D CNN, and the input shape was (3000, 1). The recordings from Sleep-EDF dataset, consisting of healthy subjects, were used as a representative to fine-tune in this experiment as it has been commonly used in EEG-only sleep staging studies. We initialized the model from the four paradigms: our proposed approach, Baseline-1, Baseline-2, and the model without the pre-training (randomly initialized using Xavier), in which each of them was performed in a total of 5 runs. As shown in \autoref{tab:results} (EEG only (1D-CNN)), while the fine-tuning using Baseline-1, Baseline-2, and our approach yielded MF1 of 50.5, 53.4, and 54.3, respectively, the model without the pre-training phase (not displayed in the \autoref{tab:results}) yielded only 18 $\pm$ 0.002. In addition, the F1-scores after training from scratch were only 17.84, 8.76, 25.27, 20.82, and 19.47 from W, N1, N2, N3, and REM, respectively. The model, not surprisingly, resulted in poor performance in line with the hypothesis that the deep neural network (DNN) would require to be trained with a large amount of data in order to achieve a high performance. Therefore, the TL paradigm could become one of the solutions for the issue of small available data.

\begin{figure*}[t]
\centering
  \includegraphics[width=\textwidth]{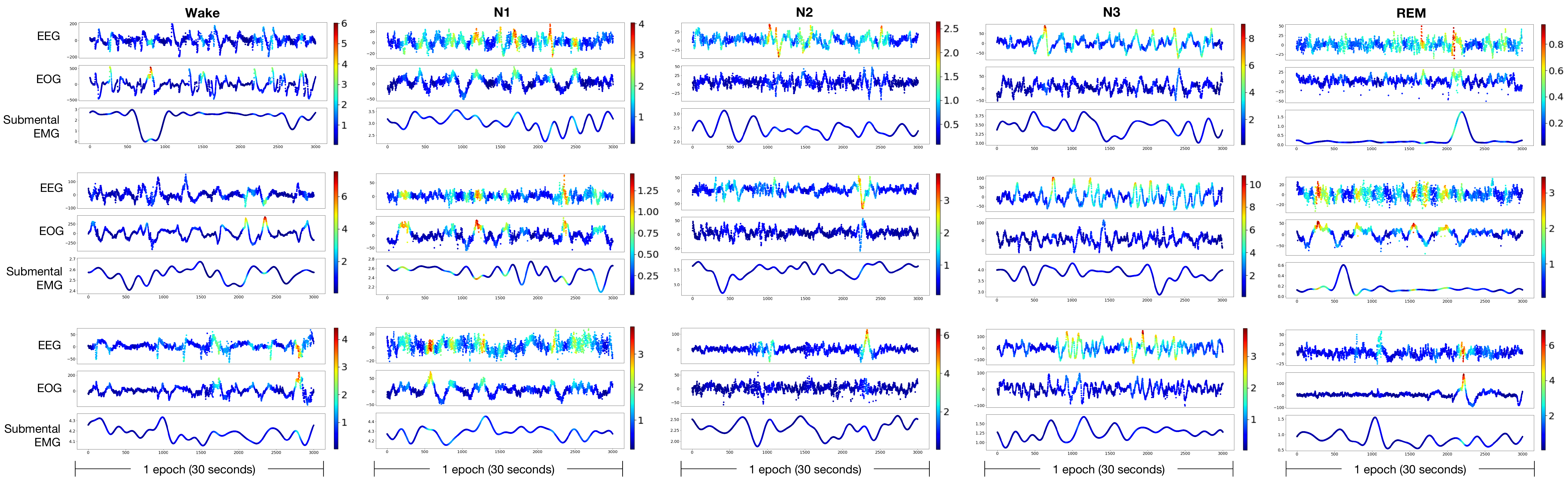}
  \caption{Visualization results from Layer-wise Relevance Propagation (LRP): Results were sampled from 2 healthy subjects in Sleep-EDF (SC). The prediction of the model on the sleep stages W, N1, N2, N3, and REM are displayed. }
  \label{fig:lrp}
\end{figure*}

% \begin{figure}[]
% \centering
%   \includegraphics[width=\linewidth]{loss.pdf}
%   \caption{Pre-training loss. Examples of training and validation loss while pre-training using the two paradigms.}
%   \label{fig:pre-train-loss}
% \end{figure}

% \begin{figure}[]
% \centering
%   \includegraphics[width=\linewidth]{loss-compare-5-folds.pdf}
%   \caption{Fine-tuning loss. Examples of fine-tuning results in which the pre-trained weights from each approach were fine-tuned separately to four subjects from different cohorts, including two healthy subjects (upper) and two patients (lower). The plots show average loss with standard deviations of 5 times running while each of them initialize with different pre-trained weights.}
%   \label{fig:loss-compare}
% \end{figure}

\subsection{Effects of input information on MetaSleepLearner performance}
\label{sec:input_vs_perf}
We hypothesized that the information of inputs might affect both performances of TL and the elicitation of our approach. The performances in the fine-tuning phase were compared among the three types of model structure and inputs: 1) 1D-CNN with EEG, 2) 2D-CNN with EEG, and 3) 2D-CNN with EEG, EOG, and sub-mental EMG. The data from Sleep-EDF (SC) were used in this experiment. As shown in \autoref{tab:results}, while using EEG only, after fine-tuning the 2D-CNN from each pre-training paradigm, all of them achieved much better performance than 1D-CNN. When the model structures were 2D-CNN, Baseline-2 surprisingly became worse while being fed with more input data. We also observed this inconsistency when experimenting on a few subjects from other databases (MASS SS2 and UCD) (not shown in the table). That is, some subjects achieved better performance with 3 modalities, while some did not. Moreover, the results from MAML were significantly better than Baseline-1. Although Cohen's kappa obtained by our approach was approximately 0.02 lower than that of Baseline-2 in case of 2D-CNN with EEG only, our approach maintained the best one in case of 3-modality input. This suggests the necessity of using all three modalities in the automatic sleep staging task in this TL setup.

% When using only EEG signals, the calculated MF1 reached only 51.91 $\pm$ 0.006 and 52.89 $\pm$ 0.006 for the conventional approach and our approach, respectively. To compare against the three modals as inputs, the network was changed to 2D CNN. The MF1 yielded from each paradigm was computed to be increased to 63.87 $\pm$ 0.007 from the conventional approach and 68.34 $\pm$ 0.006 from our approach, in which the bio-signals were from the same set of subjects within the same cohorts in only EEG condition. It is well acknowledged to conclude that EOG and sub-mental EMG are necessary for the sleep stage classification, similarly to the routine performed by clinicians. The higher MF1 within our approach suggests the necessity in using all three modals in performing the sleep stage classification and the elicitation of the performance of our approach to outperform the conventional approach ($p < 0.05$). 

\subsection{Adaptability of MetaSleepLearner in different kinds of subjects}
\label{ResultD}
Considering the advantages of MAML, our approach was expected to achieve better performance than the baselines, relying on the conventional TL approaches while enjoyed fast adaptation. In order to provide empirical evidence, we firstly inspected the cohorts of healthy subjects. The number of samples per class used for fine-tuning ($K$) was set to 5 and the maximum number of training iterations was 500. The first dataset to be explored is the ISRUC (subgroup 3) which used the same EEG electrode placements as the pre-training phase, i.e., C3-A2. The Cohen's Kappa using our approach reached 0.671 $\pm$ 0.005 while using Baseline-1 and Baseline-2 yielded only 0.633 $\pm$ 0.004 and 0.613 $\pm$ 0.005, respectively. The percentage of improvement using our approach was 6.03\%, calculated by (MAML – the best result of two baselines) / the best result of two baselines, revealing the significant difference ($p < 0.05$) in the mean of all approaches. The second dataset to be explored was CAP, despite containing the data with different EEG channels from the MASS dataset used in the pre-training. Our approach was found to statistically outperform the conventional approaches with 8.28\% of improvement in Cohen's Kappa, yielding 0.661 $\pm$ 0.006 compared to 0.611 $\pm$ 0.006 and 0.596 $\pm$ 0.006 of Baseline-1 and Baseline-2, respectively. Similarly, our approach led to an improvement of 5.4\% of Kappa on MASS SS2 over the best baseline (Baseline-1), implying the benefit of fine-tuning on the subjects whose recordings were similar to the data used in the pre-training phase. Our approach yielded 0.682 $\pm$ 0.004 while the conventional approaches produced 0.648 $\pm$ 0.004 (Baseline-1) and 0.618 $\pm$ 0.004 (Baseline-2). In a deep inspection of the results, we firstly inspected the cohorts which contained only the subjects who were labelled as healthy. The last dataset, as reported in \autoref{sec:input_vs_perf}, was Sleep-EDF (SC), showing that our approach still performed well and adapted quickly even when a different EEG electrode placements were used. An improvement of 8.48\% on accuracy over the best baseline (Baseline-2) was observed. We also showed further results of the cohort from Sleep-EDF (SC) in Appendix, in terms of confusion matrix from all subjects in the cohort and output hypnogram from subject 6. Note that the best baseline, Baseline-2, was used in these comparisons.

To determine whether our approach could be applied to a simulation of a real-world setting, the data of patients with different ratio of each sleep stage and bio-signal characteristics from the control group were used in fine-tuning. Four datasets were chosen to examine the performance of the model’s adaptation. The results on ISRUC (Subgroup 1) dataset showed that the Cohen's Kappa obtained from all pre-training paradigms was reduced compared to the control groups. Our approach achieved a Cohen's Kappa of 0.618 $\pm$ 0.002 while the Baseline-1 and Baseline-2 attained only 0.577 $\pm$ 0.002 and 0.547 $\pm$ 0.002, respectively. The approach also statistically outperformed the highest performance of conventional methods with 7.05\% Kappa improvement. The other cohorts (Sleep-EDF (ST), CAP, and UCD) were also tested. However, we found that the model performed modestly in all TF approaches, i.e., the average MF1 was less than 50. To elucidate the depleted execution, $K$ was increased to 10 epochs per sleep stage. Despite the lower performance (Cohen's Kappa of 0.476 $\pm$ 0.005, 0.471 $\pm$ 0.004, and 0.554 $\pm$ 0.004 by Baseline-1, Baseline-2, and our approach, respectively) on Sleep-EDF (ST) compared to the healthy Sleep-EDF (SC) cohort, our paradigm performed significantly progressing with the improvement of 16.2\%. Furthermore, we found that other cohorts required more training iterations. Thus, we extended the maximum of training iterations to 1000 iterations. The results on CAP dataset confirmed the increase with a Cohen's Kappa of 0.615 $\pm$ 0.004. Our approach improved 7.9\% over the conventional approach (Cohen's Kappa of 0.57 $\pm$ 0.003). Although the lowest performance was seen on UCD dataset, our approach (Cohen's Kappa of 0.429 $\pm$ 0.006) resulted in an improvement of 17.7\% on Kappa over Baseline-2 (Cohen's Kappa of 0.364 $\pm$ 0.006).

Inclusively, using the three modalities from the recordings of both healthy subjects and patients, our approach outperformed the conventional pre-training method with significant differences ($p < 0.05$) in overall accuracy, Cohen's Kappa, and MF1 in almost every sleep stage. This implied that our proposed pre-training paradigm could enhance the performance of TL while enjoying fast adaptation to the new individuals in new cohorts.

\subsection{Model Interpretation}
\label{ResultE}
In order to understand what the model learns from our approach, the evaluation by LRP was inspected, as displayed in \autoref{fig:lrp}. Three samples per sleep stage recorded from two subjects in the Sleep-EDF (SC) dataset were selected for illustration. The size of the LRP results were the same as the original inputs shape (30 seconds length $\times$ 100 sampling frequency), while the colors represented the level of effects to each prediction by the model, i.e., $R$ scores from LRP. $R$ scores were scaled with all three modalities in each sample. The blue to red colors signified the lowest to highest contribution of the prediction. Only the correct prediction samples were explored in order to determine whether the performance of the model was sufficient with correct learning method.

\autoref{fig:lrp} displays the predictions for W, N1, N2, N3, and REM stages. According to the red-colored sampling points at the left most column figures, EOG signals were found to have an impact on the prediction of W stage. This confirmed the practical views in which the EOG signals would generally show higher activity in the W stage compared to the other stages during the sleep interval. In N1 and REM stages, the similar portions of the three bio-signals were highlighted. This implied the necessity of the three modalities to assess the correct classification. In clinical settings, clinicians routinely identify these stages using all three modalities. EOG signals can be observed to be discriminative between N1 and REM as higher activity is seen in REM \cite{ronzhina2012sleep}. Similarly, the information obtained the reduced activity in sub-mental EMG signals can distinguish N1 from W stage. Furthermore, for stage N2, the model emphasized on the EEG signals, especially the area denoted with the characteristics of K-complex, which are the main EEG traits of this stage. The model also paid attention to only the EEG signals in identifying N3 stage, in which all samples in each epoch were analyzed. In accordance with its name ``Slow Wave Sleep", referring to the EEG signals with low frequency, the model required the information of the signal frequency to predict this sleep stage. To that end, the model would require the entire sample length, as illustrated with the non-dark-blue color in the visualization of the EEG modality.

The visualization results revealed that the event occurring inside each epoch displayed the most discriminative information that the model used to classify sleep stages. Moreover, the most discriminative input was found to be the samples in N2 stage, i.e., the samples could be straightforwardly classified as N2 when the K-complex was found. It also verifies that using only several epochs to fine-tune the model is possible for the model adaptation to new individuals in new cohorts as the networks are capable of learning with reasonable predictions and acceptable directions.

% \subsection{Comparison of our results to the existing work}

% \textcolor{blue}{As mentioned in \autoref{sec:motivation}, the most relevant work that aims for the same objective as our study is \cite{chen2019personalized}. They also tried to personalize sleep staging by using only a few samples. However, their proposed method, namely Symbolic Fusion, contained some limitations which can be solved by Deep Learning (DL) approach. Firstly, Their method requires domain knowledge to set-up the rules for feature extraction and sleep stage classification which can be done automatically by DL. Secondly, The decision rules were made up by domain knowledge, i.e., following the AASM, in which its advantages are reasonable and explainable. However, the rules might not be all covered because in the learning system, not only the human can teach the machine, but the machine can detect some information insight which might not be definable by humans.
%  }

\section{Discussion and Limitations}
\label{sec:discuss}
% One of the objectives of this pilot study is to explore the ability of the proposed novel pre-training method to enhance the TL approach used in sleep stage classification. Although our model was modified into a humbler version which could not be in comparison to the larger state-of-the-art classification methods, to our best knowledge, this is the first study using this task to explore a novel method of pre-training without challenging the whole process performed by the state-of-the-art models previously. Not as an alternative machinery to replace clinicians, we offer and encourage a procedure to consolidate the human-machine collaboration system for sleep stage classification with an attempt to reduce the workload and ameliorate the time-consuming task of manually labelling the sleep stages. Using the Model Agnostic Meta-Learning or MAML \cite{finn2017model}, our approach lessens the burden of the clinicians during a sleep test, labelling only 5 or 10 epochs per sleep stage in preference to one whole night while allowing the machine to handle the rest of the samples.

\begin{figure}[]
\centering
  \includegraphics[width=\linewidth]{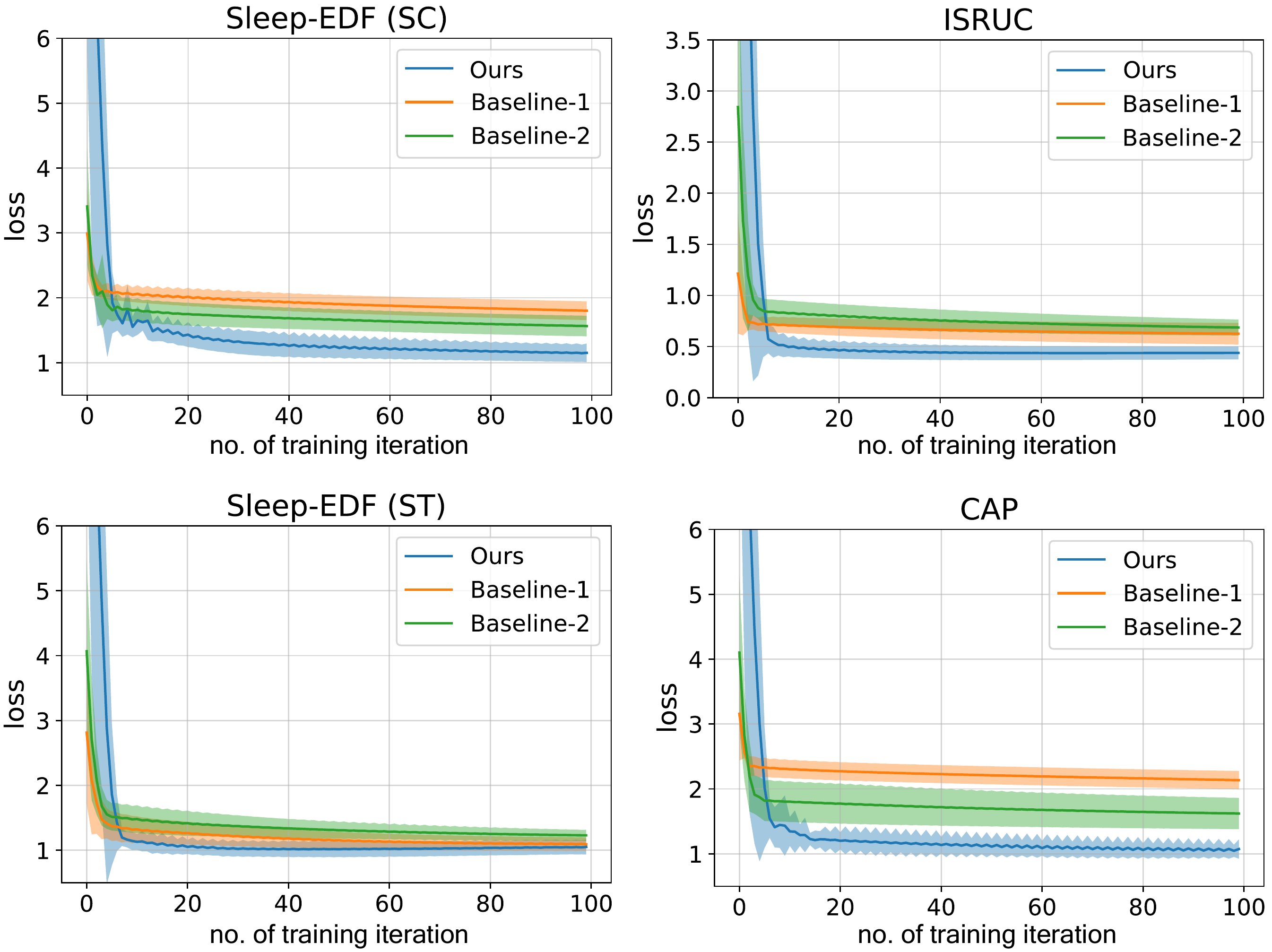}
  \caption{Fine-tuning loss. Examples of fine-tuning results in which the pre-trained weights from each approach were fine-tuned individually to two healthy subjects (upper) and two patients (lower) from four different cohorts. The plots show average loss with standard deviations of 5 times running while each of them initializes with different pre-trained weights.}
  \label{fig:loss-compare}
\end{figure}

The results confirmed the promise of our approach, outperforming the simplified version of the conventional pre-training methods in both healthy subjects and patients when the bio-signals are present with EEG, EOG, and sub-mental EMG, which are recorded generally in the medical PSG. One explanation for the achievement of the model could be attributed to the meta-learning, which minimizes the losses across all meta-tasks ($T_b$) after the adaptation of each task. The meta-weights, or $\theta_0$, were generalized while also prompting them to become more adaptable to new tasks or new individuals from newly introduced cohorts with only several samples. In contrast to the conventional pre-training approach, the model tried to minimize the loss from all samples, making the model more fitted to the training data. To support these statements, some examples of the validation loss during the fine-tuning phase are illustrated in \autoref{fig:loss-compare}. Two healthy subjects from Sleep-EDF (SC) and ISRUC as well as two patients from Sleep-EDF (ST) and CAP were selected for illustration. Only the first 100 iterations were shown, although the model was trained until fitted to the data. The middle line is the mean of the 5 runs in which each run was initialized by different pre-trained weights and fine-tuned on the selected subjects. The areas with colors represent the standard deviations (SD). It can be seen mostly in the results that during the first iteration where fine-tuning was not performed, the $\theta_0$ obtained from our approach acquired a very high loss, which was much higher than using $\theta_0$ from both Baseline-1 and Baseline-2. However, after fine-tuning was performed for only a few iterations, our proposed procedure could fast adapt to the provided data, resulting in much better and eventually higher performances were achieved, as reported in \autoref{tab:results}. Granting that the results show that there is feasibility to achieve higher performance using this approach, some limitations still remain in this pilot study, as described in the following section.

% In most of all the results, we found that the loss from our approach are usually higher than two baselines as shown in the first iteration in the plots. However, after fine-tuning for a few iterations, the loss from our approach could be noticeably observed as lower than the conventional approaches and eventually got a higher performance. 

\subsection{Adjustment of schematic frameworks}
Since our model was modified into a simpler version, to mainly explore the possibility of the proposed approach, it is incomparable to the larger state-of-the-art classification methods. Further adjustments to the base model could be explored in order to improve the effectiveness of the proposed method.

Firstly, the precision rate of sleep stage classification could be increased. The prediction of the stages N1 and REM were shown to be lower than the other stages. One reason might be due to the lower amount of N1 epochs compared to the others. However, this is to be expected as they were in accordance with the general EEG characteristics of both stages. Moreover, in the dataset consisting of the recordings from the patients, the accuracy of every stage tended to decrease. When examining the hypnogram, it is interesting to note that the surrounding epochs might affect the decision and assist in improving the performance. It could be speculated that some of the predictions could be improved when the information regarding the preceding or the succeeding epochs are presented to the model. Similarly, higher accuracy might be stemmed from an input longer than 30 seconds. Hence, the one-to-many or many-to-one training frameworks from the previous works might affect the balancing of the transitioning sleep stages \cite{sors2018convolutional, phan2019}. 

Secondly, extending the ability of the networks would unquestionably enhance the classification. In this study, we extracted a modified version of simple CNN networks, imitating from DeepSleepNet \cite{supratak2017deepsleepnet}. However, it is able to perform with any gradient-based networks. Therefore, the state-of-the-art networks such as DeepSleepNet, SeqSleepNet, or other larger networks could be applied to our proposed TL procedure in order to let the network learn contextual information between epochs using RNNs. To that extent, further exploration using another type of meta-learning which is suitable for RNNs, such as Meta-Learning with memory-augmented neural networks \cite{pmlr-v48-santoro16}, would be accommodating for this task. 

Nevertheless, the limitation of MAML involves the requirement of a large amount of computational resources due to the second-order derivation. Thus, the newer method editions, such as iMAML \cite{rajeswaran2019meta} and Reptile \cite{nichol2018first}, might help the evaluation of the larger networks. Although the performance of those methods might not be as good as the original MAML, they require lower amount of resource to operate.

\subsection{Hyperparameter K assessment}
% Both pre-training and fine-tuning are the two imperative phases in our methods, where some parameters can be tuned for an improvement.

$K$ represents the number of training samples per sleep stage. In meta-training, we appointed $K$ = {5, 10, 15} to explore the differences in which more description is available in Supplementary Materials. Not much differences between the results from different $K$ were observed. However, $K$ = 10 produced the highest performance, hence, its application in this study. In the fine-tuning phase, $K$ was set as 5 and 10 alternatively. Although these numbers were chosen in order to regulate the labelling of the samples in a new subject, it should be considered in future works as the exploring of the trade-off between the personality handling and the minimizing number of required labels from clinicians.

\subsection{Standard Hyperparameters assessment}
One of important parameters which might affect model's performance is the number of meta-tasks selected in each meta-training iteration ($\mathcal{B}$), which was currently fixed as 9 in our study due to the limitation of computation resources. The another one is the \textit{num\_updates} variable, which is the number of updating rounds inside each meta-task during meta-training. In this study, the performances of the model alternating between using 5, 10, and 15 rounds, measured by the validation loss during the pre-training process, exhibited similarities among each other. However, in some conditions, 15 rounds resulted in a higher validation loss than the other rounds. One could speculate that the higher the number of training rounds, the more general interpretation of the knowledge and the further distance from the best parameters of each meta-task ($\theta_b’$) could be achieved. Ultimately, although the result suggests that the most optimal \textit{num\_updates} for the most fitted validation loss from the three values we selected is 10 rounds, higher or lower values of \textit{num\_updates} could still be further explored.

\subsection{Benefits from actual practicality and future works}
In order to drive further impact stemmed from our objective of enhancing the human-machine collaboration, it is advantageous for the model to guide the clinicians of which epochs should be labelled, leading to an effective fine-tuning of the model. One approach that could be implemented to enhance the predictive performance is called active learning \cite{activelearning}. It points out which samples, i.e., epochs, benefit the model, in which low confidence in prediction might be exhibited, and presents them to the clinicians to provide their expertise and label the samples efficiently \cite{liang2019development}. 

An alternative way to use learning feedback for improvement is to perform model interpretation. Any improvements will be valuable if the model presents the results and learns in the most possible factual way. Since it is challenging to comprehend the learning contents of the DNNs despite the application of TL, an effectual feedback system is crucial to review the results. The utilization of Layer-wise Relevance Propagation or LRP ensures that the decisions generated by the models stemmed from the same reason the clinicians might use for annotation. One possibility involves the performance metrics showing that the prediction of N1 could be mistakenly classified as REM and vice versa. The visualization states for the reason that it is not only because of the similarity in the characteristics of EEG signals, but also the ways the model learns in both classes. However, our results suggest that the model did not learn to make a reasonably decision in some patients, leading to a lower quality performance in the patients in comparison to the healthy controls. Therefore, the concern over the interpretation of any models along with the adjustment of the other networks or training procedures should be investigated to ensure the precision of the improvement.

The filtering of the subjects with insufficient samples per sleep stage also affected the learning ability of the model. For illustration, we separated those who held samples less than $K \times 3$ epochs per sleep stage to another table, shown in \autoref{tab:results2} in Appendix. The samples from each subject were selected in the same paradigm as our main experiments, which were selecting $K$ samples per sleep stage to train, another $K$ to validate, and the rest for testing. Therefore, if any subjects contained only a few samples, i.e., less than $K \times 3$, there would be only a few samples left for testing. Additionally, if the number of samples were less than $K \times 2$, the samples would not be enough for validation. However, our proposed approach still mostly outperformed both of the conventional baselines with significant differences (p $<$ 0.05).

In addition, the pre-training procedures can be improved by feeding more datasets to facilitate the learning of the pre-trained models from subjects with different demographic and clinical backgrounds. Despite its advantage, data variation from different cohorts may complicate the model processing. Thus, a deep inspection is essential. Moreover, the data from an actual clinical oriented recording (non-public) may have proven important for data diversity. However, more datasets are typically accompanied by longer computational time and more efforts for pre-training. In order to mitigate the problem, the clustering method for grouping similar subjects and the representatives selection instead of using every subject are appealing approaches to be applied in the future.
% Currently, our team at the Vidyasirimedhi Institute of Science \& Technology (VISTEC) in collaboration with the Center of Excellence in Sleep Disorders, King Chulalongkorn Memorial Hospital, Bangkok, Thailand, has been establishing a partnership, aiming to assemble and collect various sleep data in patients during an actual PSG for future development of sleep lab technology.

% The filtering of the subjects with insufficient samples per sleep stage also affected the learning ability of the model. Albeit a low number of filtered out subjects, it implies data manipulation, requiring a necessity to search for a new solution in order to serve in an actual PSG. In terms of the bio-signal data, only EEG channels selected in this study were bipolar electrodes, commonly used in all datasets. It would be beneficial to explore other various EEG channels from the selected datasets, allowing an extensive usage of data in the new cohorts. 

\section{Conclusion}
This pilot study explored a feasibility to perform fast adaptation by applying a Model Agnostic Meta-Learning (MAML) approach, in order to transfer the acquired sleep staging knowledge from a large dataset to new individuals in new cohorts. A simplified edition of the Convolutional Neural Network (CNN) was pre-trained using our approach with MASS dataset, followed by the adaptation or the fine-tuning to each new subject from other cohorts, including Sleep-EDF, CAP, ISRUC, and UCD, by using only several samples from each subject. The performance was compared against the pre-training of two conventional approaches. The investigation using only EEG signals confirmed that only one modality of recordings contained insufficient knowledge for sleep stage classification. Subsequently, three bio-signal modalities (EEG, EOG, and sub-mental EMG) were employed. One conventional approach displayed poor performance when it retrieved more modalities of input, while both our approach and the second conventional approach achieved higher performance. Moreover, our approach statistically outperformed the conventional approaches as a result from the 5 runs of pre-training and 5 runs of fine-tuning to each individual using the recordings of both healthy subjects and patients. The study also illustrated the learning of the model after the adaptation to each subject, ensuring that the performance was directed towards reasonable learning. This indicates a possibility of using this framework for human-machine collaboration for sleep stage classification, easing the burden of the clinicians in labelling the sleep stages through only several epochs rather than an entire recording.

\bibliography{References}
\bibliographystyle{IEEEtran}

\newpage

\begin{appendices}

\section{}
\label{app:data}

Five publicly datasets were used to evaluate our method. The number of subjects and EEG electrode placements used in each dataset are explained in the \autoref{tab:results} and \autoref{tab:results2}. The study was approved by Rayong Hospital Research Ethics Committee (RYH REC No.E008/2562), Thailand.

\subsection{Montreal Archive of Sleep Studies (MASS)}
MASS is a collection of a total of 200 polysomnograms (PSG) from 97 male and 103 female subjects \cite{mass}. The cohort was split into five subsets (SS1-SS5), according to the research protocols used when collecting the data. These recordings were manually classified by sleep experts using AASM guidelines with 30-second epoch length for SS1 and SS3 and R\&K rules with 20-second epoch length for SS2, SS4, and SS5. The labels following the classification by R\&K comprised eight classes, namely sleep stages W, S1, S2, S3, S4, REM, and MT, with those that are unable to be classified labelled as ``UNKNOWN” \cite{moser2009sleep, phan2019}. In this study, the AASM guidelines with five sleep stages were followed instead for consistency, merging sleep stages S3 and S4 into N3 and the segments of MT and ``UNKNOWN” were removed. Furthermore, the EEG and EOG recordings were pre-processed with a notch filter of 60 Hz and band-pass filters of 0.3 - 35 Hz. All recordings from the original sampling frequency of 256 Hz were downsampled to 100 Hz. All segments were generated into 30-second long, extending both ends by 5 seconds for segments that were originally 20-second long.

\subsection{Sleep-EDF}
The Sleep-EDF dataset \cite{kemp2000analysis} contains two sets of data from two studies: the effects of age on sleep in healthy controls (SC) and the effects of temazepam, a benzodiazepine, on sleep in subjects with difficulty falling asleep (ST). For SC, we used only the same set of subjects as DeepSleepNet \cite{supratak2017deepsleepnet}, such that the number of subjects from SC are 20. For ST, we used all subjects provided from the dataset, which are 22. These recordings had a sampling rate of 100 Hz and were classified according to R\&K rules, which were merged into five sleep stages following the AASM standard to be consistent with MASS. Due to the long periods of W stage at the beginning and at the end on most of the recordings, the methods were also modified following DeepSleepNet, by truncating the awake periods at each end to at most 30 minutes. In addition, most subjects contain two-night recorded signals, but only the data obtained during the first night from each subject were used in our experiment.

\subsection{CAP Sleep Database}
The title of the Cyclic Alternating Pattern (CAP) Sleep Database is originated from the recurring EEG activity at intervals during NREM \cite{terzano2001atlas}, where its irregularity may imply a range of sleep disorders. The database consists of 108 PSG recordings, including 16 recordings from healthy subjects and 92 pathological recordings, which were categorized into nocturnal frontal lobe epilepsy (40), REM behavior disorder (22), periodic leg movements (10), insomnia (9), narcolepsy (5), sleep-disordered breathing (4), and bruxism (2). However, we found a problem in accessing the data from 1 bruxism subject. Thus, the remaining 91 pathological recordings were used in our experiment. The sleep stages were scored by expert neurologists according to the R\&K rules \cite{goldberger2000physiobank}, but also modified according to the AASM standard, similarly to the other datasets.

\subsection{ISRUC}
A total of 118 PSG recordings, named ISRUC-Sleep dataset \cite{isruc}, was introduced in 2015 by a team at the Sleep Medicine Centre of the Hospital of Coimbra University (CHUC). The dataset includes three separated subgroups of PSG signals from 100 subjects with history of sleep disorders recorded on one data acquisition session, 8 subjects recorded on two different sessions and two different dates, and 10 healthy subjects. We selected only subgroup 1 and 3 to be representatives in our experiment. The sleep stage classification followed the five sleep stages according to the AASM, labelled by two human experts in each session. 

\subsection{St. Vincent's University Hospital / University College Dublin Sleep Apnea Database (UCD)}
Revised in 2011, the St. Vincent's University Hospital / University College Dublin Sleep Apnea Database, abbreviated as UCD \cite{goldberger2000physiobank}, holds the overnight PSG of 25 subjects (21 male and 4 female) with diagnoses of possible sleep-related breathing disorders such as OSA, central sleep apnea, and snoring. Sleep stages were scored by sleep experts, following the R\&K rules, into 8 stages: W, S1, S2, S3, S4, REM, Artifact, and Indeterminate. Similar to other datasets, S3 and S4 were merged into one stage in order to comply with AASM standard and the segments Artifact and Indeterminate were removed in this study. The recordings were pre-processed with a notch filter of 60 Hz, followed by the downsampling from the original sampling frequency of 128 Hz to 100 Hz for consistency with the other datasets. The W periods at the beginning and at the end of each recording which were longer than 30 minutes were trimmed to approximately 30 minutes on both ends.

\section{}
The meta-training procedure, which is described in \autoref{sec:fast-adapt}, could be illustrated as follows:
\label{app:meta-train}
\begin{figure}[h]
\centering
  \includegraphics[width=\linewidth]{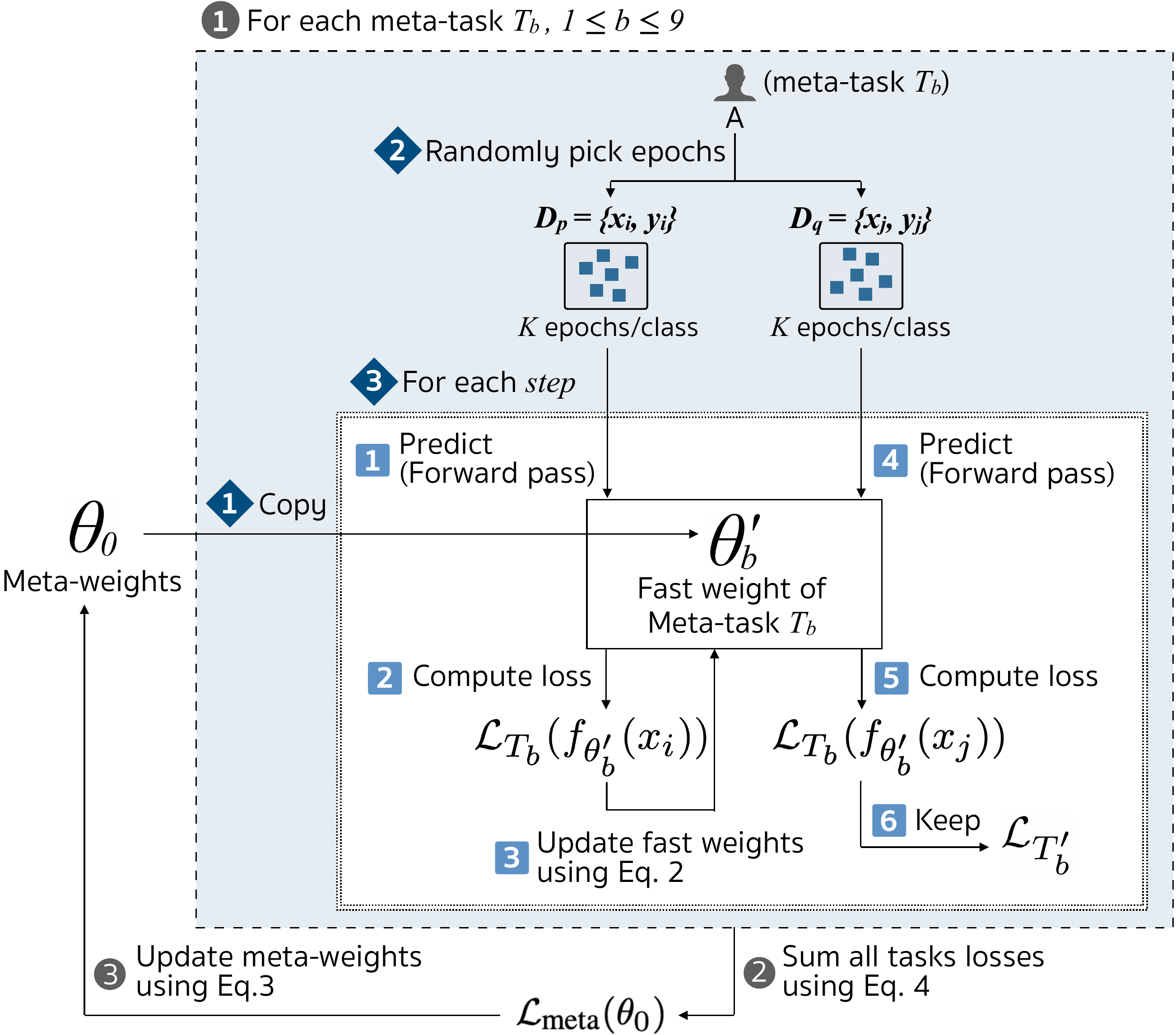}
  \caption{Diagram of one meta-training iteration.}
  \label{fig:meta-train}
\end{figure}

\section{}
\label{app:model}
\begin{figure}[h]
\centering
  \includegraphics[width=\linewidth]{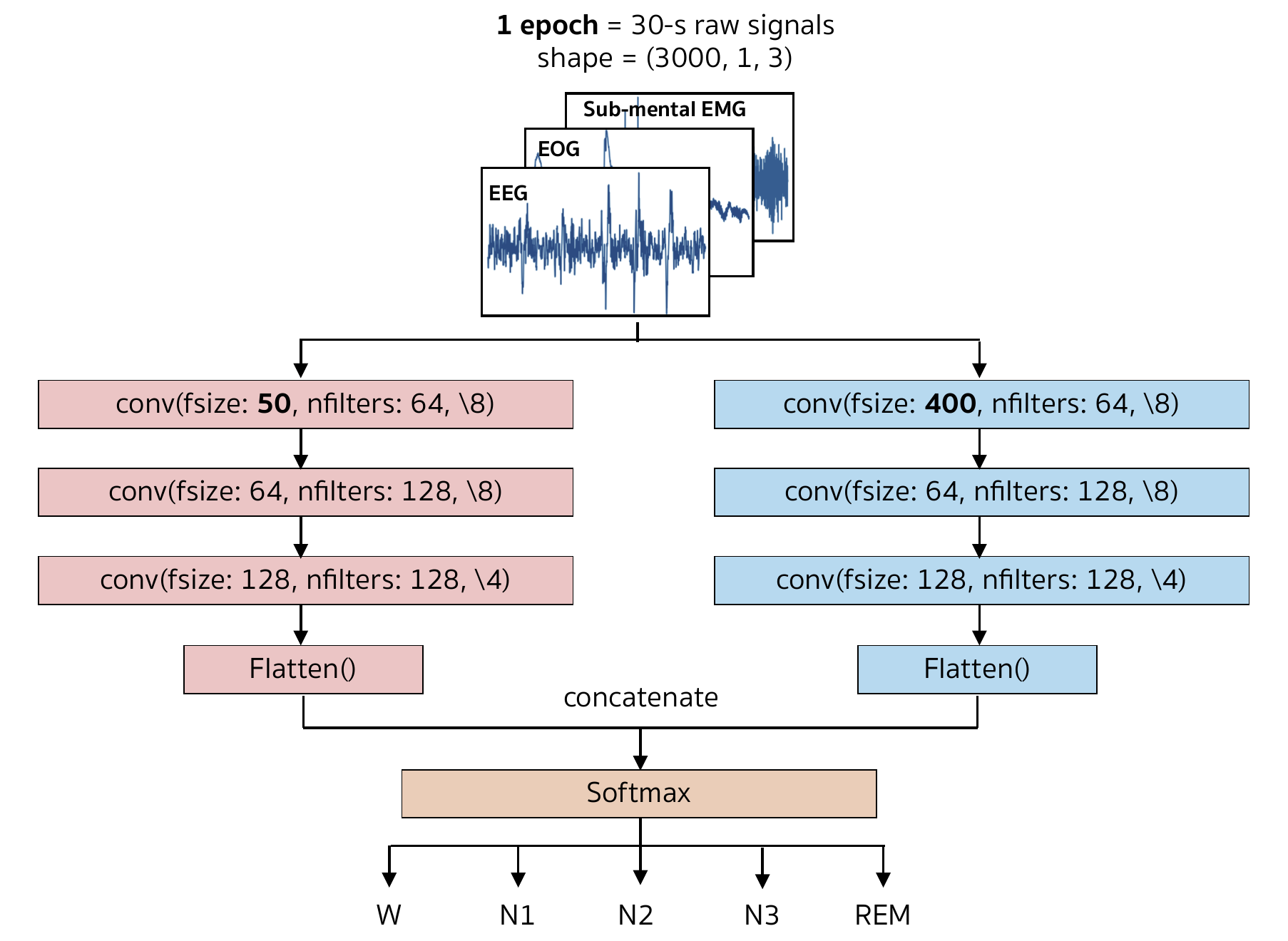}
  \caption{Example of sleep stage classification network, demonstrating the performance of our proposed fast adaptation procedure. 1 epoch (1 input sample) contains 30 seconds of EEG, EOG, and sub-mental EMG signals. The conv blocks refer to the CNN layers, each with a variety of filter size, number of filters, and stride size, respectively. The bold numbers in the first conv blocks of each side represent the small (left) and large (right) filters of the CNN layers.}
  \label{fig:model}
\end{figure}

The model, shown in \autoref{fig:model}, designed as a simpler version of DeepSleepNet, composes of two stacks of CNN layers: small filters (the left/pink stack) and large filters (the right/blue stack), in order to capture the temporal and frequency information, respectively. Each CNN layer is followed by the $relu$ activation function \cite{relu}. The \textit{l2} regularization is used in the first CNN layers as the original network, in order to prevent the over-fitting of noises and artifacts from the signals. However, all CNN layers were changed from 1D to 2D, resembling the study by Phan \textit{et al.} \cite{phan2019towards} to support multi-modal signals.

For the input data, as described in Appendix \ref{app:data}, only MASS dataset was bandpass filtered with the proper frequency range in order to be used as a pre-training data. For other cohorts, all signals were notch-filtered with either 50 or 60 Hz, depending on the datasets and were re-sampled data to reach 100 Hz sampling frequency (if necessary), without any further pre-processing. We assumed that the procedure could handle the raw data from different hardware devices along with different pre-processing methods such as hardware filters.

The model was implemented using \textit{Tensorflow 1.13.1} \cite{tensorflow2015}. The required inputs in shapes $(width, height, number\_of\_channels)$ could be described as $(no\_of\_samplings, 1, no\_of\_modals)$ = (3000, 1, 3), as the inputs were raw signals from the three modalities. After the inputs were passed through the CNN layers from both large and small filters concurrently, the features extracted from both sides were flatten and concatenated. At the end of the process, one Fully Connected (FC) layer with Softmax activation was used for sleep stage classification. The network was trained using Adam optimizer \cite{adam}, to minimize the cross entropy loss:

\begin{equation}
  \label{eq:loss}
  \mathcal{L}_{T_b}(f_{\theta}) = -\frac{1}{N}\left(\sum_{i=1}^{N} y_i \cdot log(f_{\theta}) \right),
\end{equation}

where $y_i$ is ground truth of sample $i$ and N is number of samples.

\onecolumn
% \section{}
\label{app:conf}
\begin{figure*}[]
\centering
%   \vspace*{-0.4in}
  \includegraphics[width=\textwidth]{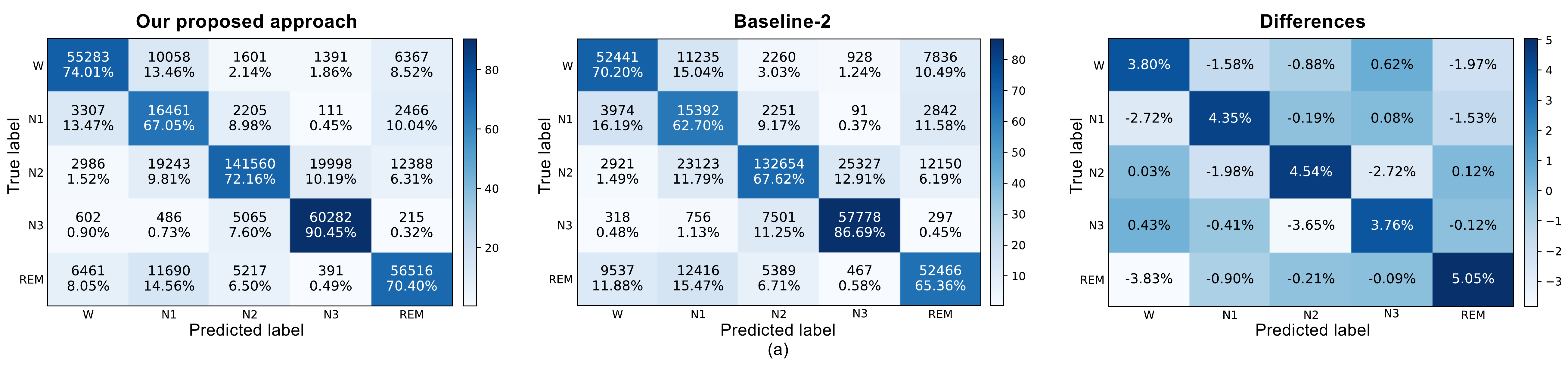}
  \label{fig:confsc}
\end{figure*}

\begin{figure*}[]
\centering
  \vspace*{-0.2in}
  \includegraphics[width=\textwidth]{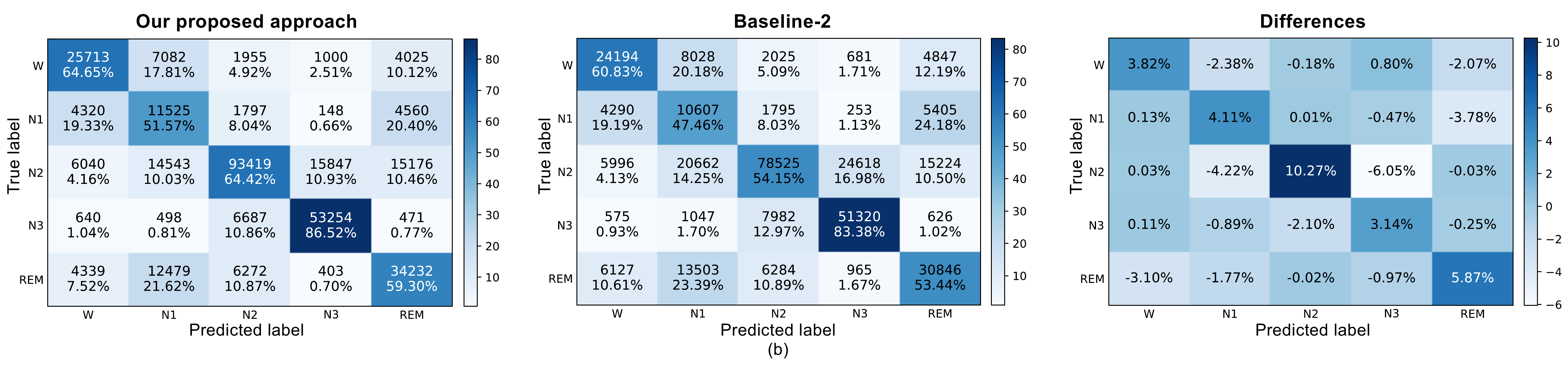}
  \caption{Examples of confusion matrix after the fine-tuning from pre-trained weights of the proposed approach (left), the Baseline-2 (center), and the differences between both results (right) using (a) Sleep-EDF (SC - Healthy Subjects) and (b) Sleep-EDF (ST - Patients). The results are the summation from all subjects and all runs.}
  \label{fig:confst}
\end{figure*}

\begin{figure*}[]
\centering
  \vspace*{-0.3in}
  \includegraphics[width=\textwidth]{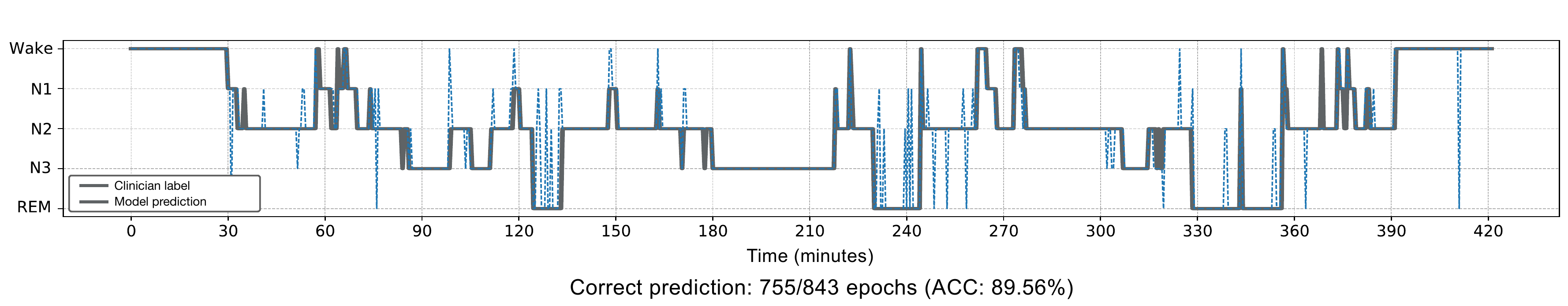}
  \caption{The hypnogram displaying the comparison between the labelling by clinician and the model's prediction of a representative subject from Sleep-EDF (SC).}
  \label{fig:hypnogram}
\end{figure*}

\begin{table*}[b]
\centering
\fontsize{19}{24}\selectfont 
\caption{Performance of five sleep stages classification after fine-tuning on each individual in each cohort. The results are averaged from all 5 pre-trained weights per paradigm $\times$ no. of subjects $\times$ 5 times of random samples selection. The subjects reported in this table are only those whose samples are less than $K \times 3$.}
\label{tab:results2}
\resizebox{\textwidth}{!}{%
\begin{threeparttable}
\begin{tabular}{|l|c|c|rrr|rrr|rrr|rrr|rrr|rrr|rrr|rrr|}
\hline
\multicolumn{1}{|c|}{\multirow{3}{*}{\textbf{Dataset}}} & \multirow{3}{*}{\textbf{\begin{tabular}[c]{@{}c@{}}EEG\\ channel\end{tabular}}} & \multirow{3}{*}{\textbf{\begin{tabular}[c]{@{}c@{}}No. of\\ subjects\end{tabular}}} & \multicolumn{3}{c|}{\multirow{2}{*}{\textbf{\begin{tabular}[c]{@{}c@{}}Overall \\ Accuracy (\%)\end{tabular}}}} & \multicolumn{15}{c|}{\textbf{F1 per class}} & \multicolumn{3}{c|}{\multirow{2}{*}{\textbf{MF1}}} & \multicolumn{3}{c|}{\multirow{2}{*}{\textbf{Cohen's kappa ($\mathcal{K}$)}}} \\ \cline{7-21}
\multicolumn{1}{|c|}{} &  &  & \multicolumn{3}{c|}{} & \multicolumn{3}{c|}{\textbf{W}} & \multicolumn{3}{c|}{\textbf{N1}} & \multicolumn{3}{c|}{\textbf{N2}} & \multicolumn{3}{c|}{\textbf{N3}} & \multicolumn{3}{c|}{\textbf{REM}} & \multicolumn{3}{c|}{} & \multicolumn{3}{c|}{} \\
\multicolumn{1}{|c|}{} &  &  & \multicolumn{1}{c}{\textbf{B-1}} & \multicolumn{1}{c}{\textbf{B-2}} & \multicolumn{1}{c|}{\textbf{ours}} & \multicolumn{1}{c}{\textbf{B-1}} & \multicolumn{1}{c}{\textbf{B-2}} & \multicolumn{1}{c|}{\textbf{ours}} & \multicolumn{1}{c}{\textbf{B-1}} & \multicolumn{1}{c}{\textbf{B-2}} & \multicolumn{1}{c|}{\textbf{ours}} & \multicolumn{1}{c}{\textbf{B-1}} & \multicolumn{1}{c}{\textbf{B-2}} & \multicolumn{1}{c|}{\textbf{ours}} & \multicolumn{1}{c}{\textbf{B-1}} & \multicolumn{1}{c}{\textbf{B-2}} & \multicolumn{1}{c|}{\textbf{ours}} & \multicolumn{1}{c}{\textbf{B-1}} & \multicolumn{1}{c}{\textbf{B-2}} & \multicolumn{1}{c|}{\textbf{ours}} & \multicolumn{1}{c}{\textbf{B-1}} & \multicolumn{1}{c}{\textbf{B-2}} & \multicolumn{1}{c|}{\textbf{ours}} & \multicolumn{1}{c}{\textbf{B-1}} & \multicolumn{1}{c}{\textbf{B-2}} & \multicolumn{1}{c|}{\textbf{ours}} \\ \hline
\textbf{Healthy} & \multicolumn{1}{l|}{} & \multicolumn{1}{l|}{} & \multicolumn{1}{l}{} & \multicolumn{1}{l}{} & \multicolumn{1}{l|}{} & \multicolumn{1}{l}{} & \multicolumn{1}{l}{} & \multicolumn{1}{l|}{} & \multicolumn{1}{l}{} & \multicolumn{1}{l}{} & \multicolumn{1}{l|}{} & \multicolumn{1}{l}{} & \multicolumn{1}{l}{} & \multicolumn{1}{l|}{} & \multicolumn{1}{l}{} & \multicolumn{1}{l}{} & \multicolumn{1}{l|}{} & \multicolumn{1}{l}{} & \multicolumn{1}{l}{} & \multicolumn{1}{l|}{} & \multicolumn{1}{l}{} & \multicolumn{1}{l}{} & \multicolumn{1}{l|}{} & \multicolumn{1}{l}{} & \multicolumn{1}{l}{} & \multicolumn{1}{l|}{} \\ \hline
\textbf{2D-CNN} &  & \multicolumn{1}{l|}{} & \multicolumn{1}{l}{} & \multicolumn{1}{l}{} & \multicolumn{1}{l|}{} & \multicolumn{1}{l}{} & \multicolumn{1}{l}{} & \multicolumn{1}{l|}{} & \multicolumn{1}{l}{} & \multicolumn{1}{l}{} & \multicolumn{1}{l|}{} & \multicolumn{1}{l}{} & \multicolumn{1}{l}{} & \multicolumn{1}{l|}{} & \multicolumn{1}{l}{} & \multicolumn{1}{l}{} & \multicolumn{1}{l|}{} & \multicolumn{1}{l}{} & \multicolumn{1}{l}{} & \multicolumn{1}{l|}{} & \multicolumn{1}{l}{} & \multicolumn{1}{l}{} & \multicolumn{1}{l|}{} & \multicolumn{1}{l}{} & \multicolumn{1}{l}{} & \multicolumn{1}{l|}{} \\
\textbf{EEG, EOG, Submental EMG} &  & \multicolumn{1}{l|}{} & \multicolumn{1}{l}{} & \multicolumn{1}{l}{} & \multicolumn{1}{l|}{} & \multicolumn{1}{l}{} & \multicolumn{1}{l}{} & \multicolumn{1}{l|}{} & \multicolumn{1}{l}{} & \multicolumn{1}{l}{} & \multicolumn{1}{l|}{} & \multicolumn{1}{l}{} & \multicolumn{1}{l}{} & \multicolumn{1}{l|}{} & \multicolumn{1}{l}{} & \multicolumn{1}{l}{} & \multicolumn{1}{l|}{} & \multicolumn{1}{l}{} & \multicolumn{1}{l}{} & \multicolumn{1}{l|}{} & \multicolumn{1}{l}{} & \multicolumn{1}{l}{} & \multicolumn{1}{l|}{} & \multicolumn{1}{l}{} & \multicolumn{1}{l}{} & \multicolumn{1}{l|}{} \\
Sleep-EDF (SC) & Fpz-Cz & 2 & 57.4 & 54.3 & 58.6 & 65.8 & 61.8 & \textbf{62.4} & 17.8 & 15.0 & 18.7 & 65.2 & 63.1 & 67.3 & 55.5 & 53.0 & 55.6 & 57.1 & 52.1 & 56.4 & 52.2 & 48.9 & 51.9 & 0.407 & 0.368 & 0.415 \\
CAP & \begin{tabular}[t]{@{}c@{}}C3-A2 /\\ C3-P3\end{tabular} & 7 & 73.5 & 69.8 & \textbf{75.4} & 67.5 & 61.6 & \textbf{71.8} & 11.4 & 10.4 & \textbf{13.5} & 71.6 & 68.7 & 73.1 & 85.3 & 81.6 & 85.5 & 71.9 & 68.0 & \textbf{74.3} & 66.8 & 63.2 & \textbf{68.8} & 0.630 & 0.580 & \textbf{0.654} \\ \hline
\textbf{Patients} &  &  & \multicolumn{1}{c}{} & \multicolumn{1}{c}{} & \multicolumn{1}{c|}{} & \multicolumn{1}{c}{} & \multicolumn{1}{c}{} & \multicolumn{1}{c|}{} & \multicolumn{1}{c}{} & \multicolumn{1}{c}{} & \multicolumn{1}{c|}{} & \multicolumn{1}{c}{} & \multicolumn{1}{c}{} & \multicolumn{1}{c|}{} & \multicolumn{1}{c}{} & \multicolumn{1}{c}{} & \multicolumn{1}{c|}{} & \multicolumn{1}{c}{} & \multicolumn{1}{c}{} & \multicolumn{1}{c|}{} & \multicolumn{1}{c}{} & \multicolumn{1}{c}{} & \multicolumn{1}{c|}{} & \multicolumn{1}{c}{} & \multicolumn{1}{c}{} & \multicolumn{1}{c|}{} \\ \hline
\textbf{2D-CNN} &  & \multicolumn{1}{l|}{} & \multicolumn{1}{l}{} & \multicolumn{1}{l}{} & \multicolumn{1}{l|}{} & \multicolumn{1}{l}{} & \multicolumn{1}{l}{} & \multicolumn{1}{l|}{} & \multicolumn{1}{l}{} & \multicolumn{1}{l}{} & \multicolumn{1}{l|}{} & \multicolumn{1}{l}{} & \multicolumn{1}{l}{} & \multicolumn{1}{l|}{} & \multicolumn{1}{l}{} & \multicolumn{1}{l}{} & \multicolumn{1}{l|}{} & \multicolumn{1}{l}{} & \multicolumn{1}{l}{} & \multicolumn{1}{l|}{} & \multicolumn{1}{l}{} & \multicolumn{1}{l}{} & \multicolumn{1}{l|}{} & \multicolumn{1}{l}{} & \multicolumn{1}{l}{} & \multicolumn{1}{l|}{} \\
\textbf{EEG, EOG, Submental EMG} &  & \multicolumn{1}{l|}{} & \multicolumn{1}{l}{} & \multicolumn{1}{l}{} & \multicolumn{1}{l|}{} & \multicolumn{1}{l}{} & \multicolumn{1}{l}{} & \multicolumn{1}{l|}{} & \multicolumn{1}{l}{} & \multicolumn{1}{l}{} & \multicolumn{1}{l|}{} & \multicolumn{1}{l}{} & \multicolumn{1}{l}{} & \multicolumn{1}{l|}{} & \multicolumn{1}{l}{} & \multicolumn{1}{l}{} & \multicolumn{1}{l|}{} & \multicolumn{1}{l}{} & \multicolumn{1}{l}{} & \multicolumn{1}{l|}{} & \multicolumn{1}{l}{} & \multicolumn{1}{l}{} & \multicolumn{1}{l|}{} & \multicolumn{1}{l}{} & \multicolumn{1}{l}{} & \multicolumn{1}{l|}{} \\
Sleep-EDF (ST) & Fpz-Cz & 7 & 60.1 & 58.4 & \textbf{64.9} & 35.1 & 35.5 & \textbf{38.2} & 24.9 & 23.8 & \textbf{29.7} & 72.8 & 70.5 & \textbf{77.1} & 52.5 & 50.5 & \textbf{54.8} & 55.9 & 57.5 & \textbf{61.3} & 49.2 & 48.5 & \textbf{53.4} & 0.416 & 0.401 & \textbf{0.473} \\
ISRUC (Subgroup 1) & \begin{tabular}[t]{@{}c@{}}C3-A2 /\\ C3-M2\end{tabular} & 4 & 76.0 & 75.5 & \textbf{79.9} & 84.5 & 84.8 & \textbf{89.4} & 26.4 & 25.9 & \textbf{31.7} & 63.6 & 59.8 & 64.9 & 84.9 & 84.1 & 85.8 & - & - & - & 64.9 & 63.7 & \textbf{68.0} & 0.629 & 0.617 & \textbf{0.685} \\
UCD & C3-A2 & 3 & 48.3 & 48.8 & \textbf{53.1} & 54.7 & 58.6 & \textbf{64.6} & 34.1 & 31.4 & 32.5 & 43.1 & 44.1 & \textbf{51.4} & 85.2 & 91.1 & \textbf{93.8} & 33.5 & 34.6 & \textbf{44.6} & 45.7 & 46.9 & \textbf{52.4} & 0.313 & 0.329 & \textbf{0.384} \\
CAP (Patients) & \begin{tabular}[t]{@{}c@{}}C3-A2 /\\ C3-P3 /\\ C4-A1\end{tabular} & 42 & 68.0 & 67.2 & \textbf{70.5} & 66.0 & 65.6 & \textbf{71.1} & 12.0 & 11.4 & \textbf{13.4} & 65.0 & 63.4 & \textbf{66.9} & 75.7 & 74.6 & \textbf{76.4} & 63.7 & 63.8 & \textbf{67.3} & 63.4 & 62.7 & \textbf{66.1} & 0.554 & 0.543 & \textbf{0.586} \\ \hline
\end{tabular}%

% \caption*{\footnotesize The \textbf{bold} numbers represent the highest performance among all paradigms with significant difference from others (p $<$ 0.05). \\ B-1 = Baseline-1, B-2 = Baseline-2, ours = our proposed approach \\ 3 modals = EEG, EOG, and submental EMG \\ ``-" (hyphen) means all 4 subjects from ISRUC (Subgroup 1) don't have any REM samples left for testing.}
\begin{tablenotes}[para,flushleft]
The \textbf{bold} numbers represent the highest performance among all paradigms with significant difference (p $<$ 0.05). \\ B-1 = Baseline-1, B-2 = Baseline-2, ours = our proposed approach \\ ``-" (hyphen) designates all 4 subjects from ISRUC (Subgroup 1), which no REM sample remained for testing.
  \end{tablenotes}
  
\end{threeparttable}
}
\end{table*}

\end{appendices}

\end{document}